\documentclass{IEEEtran}
\usepackage{cite}
\usepackage{amsmath,amssymb,amsfonts}
\usepackage{graphicx}
\usepackage{color}
\usepackage{multirow}
\usepackage{makecell}

\PassOptionsToPackage{hyphens}
{url}\usepackage{hyperref}
\usepackage{textcomp,nicefrac}
\def\BibTeX{{\rm B\kern-.05em{\sc i\kern-.025em b}\kern-.08em
T\kern-.1667em\lower.7ex\hbox{E}\kern-.125emX}}
\begin{document}
\title{Development and Flight Trial of a UAV-based Gamma Ray and Neutron Detection System for Large-Area Radioactivity Mapping and Source Activity Estimation}
\author{Lysander Miller, Airlie Chapman, James Kennedy, Richard Hebden, and Jeremy M.\ C.\ Brown
\thanks{This work was supported by an Australian Government Research Training Program (RTP) Scholarship and DMTC Ltd. [Project: 15.188 UAV-mounted Scintillator Sensor for Standoff Detection of Radiological and Nuclear (RN) materials]. The authors have prepared this paper in accordance with the intellectual property rights granted to partners from the original DMTC project.}
\thanks{Lysander Miller and Airlie Chapman are with the Department of Mechanical Engineering, The University of Melbourne, Parkville, VIC 3010, Australia (e-mail: lysanderm@student.unimelb.edu.au; airlie.chapman@unimelb.edu.au).}
\thanks{James Kennedy is with the Defence Science and Technology Group, Department of Defence, Australia.}
\thanks{Richard Hebden is with Swinburne University of Technology, Hawthorn, VIC 3122, Australia.}
\thanks{Jeremy M. C. Brown is with the Optical Sciences Centre, Department of Physics and Astronomy, Swinburne University of Technology, Hawthorn, VIC 3122, Australia (e-mail: jmbrown@swin.edu.au).}}

\maketitle

\begin{abstract}
Advances in scintillation crystal and Silicon PhotoMultiplier (SiPM) technologies have enabled the development of compact, lightweight, and low-power radiation detectors that are suitable for integration with Unmanned Aerial Vehicles (UAVs). This integration enables efficient and cost-effective large-area radiation monitoring while minimising occupational exposure. In this work, a SiPM-based NaIL scintillation detection payload was developed, characterised, and mounted on a multirotor UAV for gamma ray and neutron source localisation and activity estimation applications. To support these capabilities, an analytic radionuclide detection efficiency model was developed and used to estimate radioactivity on the ground from aerial energy spectrum measurements. The analytic expression for the detection efficiency incorporated physical phenomena, including the branching ratio, detector solid angle, air attenuation, and intrinsic peak efficiency, leading to agreement within 10\% of experimental radionuclide detection efficiencies. The UAV-based radiation detection system was physically validated through a controlled indoor live radioactive source demonstration at 1.5 m, 3 m, and 4.5 m flight heights. Using the developed ground-level radioactivity estimation method, $^{137}$Cs and $^{60}$Co sources were successfully localised within 0.5 m, and their activities were estimated with errors on the order of 10\% or less.
\end{abstract}


\section{Introduction}\label{sec:introduction}
Radiation detection is essential for homeland security \cite{Vetter2018, Hamrashdi2019}, nuclear decontamination \cite{Sanada2015, MacFarlane2014}, and the identification of Naturally Occurring Radioactive Material (NORM) \cite{Marques2021}. These scenarios often involve large areas, obstructed terrain, and hazardous conditions that make conventional handheld detectors impractical. Radiation detectors integrated with Unmanned Aerial Vehicles (UAVs) address these challenges by enabling large-area monitoring over complex environments while eliminating the need for occupational exposure. Notable examples of UAV-based radiation detection include the unmanned helicopter mapping the Fukushima Daiichi nuclear power plant \cite{Sanada2015}, multirotor screening shipping containers for illicit radioactive materials \cite{Marques2022}, and fixed-wing aircraft surveying the Chernobyl Exclusion Zone \cite{Connor2020}. While unmanned helicopter-based systems offer high sensitivity by accommodating large-volume radiation detectors, they tend to be heavier and more expensive than fixed-wing or multirotor alternatives \cite{Marques2021}. Multirotor UAVs can provide higher spatial resolution compared to fixed-wing UAVs because of their ability to hover during measurement \cite{Ardiny2024}.

Silicon PhotoMultiplier (SiPM)-based scintillation detectors are commonly integrated with multirotor UAVs for gamma ray radiation monitoring and source localisation \cite{Ardiny2024, Marques2022, Mascarich2018, Vetter2019}. These detectors typically consist of a scintillation crystal optically bonded to a SiPM array. Gamma rays interact with the scintillator primarily through photoelectric absorption, Compton scattering, and pair production, resulting in the partial or complete transfer of its energy to the electrons in the crystal lattice. These excited electrons de-excite to the valence band via the emission of optical photons from the scintillation material. The number of optical photons produced per MeV of deposited gamma ray energy (known as the optical yield) is characteristic of the scintillator. The optical photons interact with the SiPM array, producing an analog pulse with amplitude proportional to the number of optical photons detected \cite{Knoll2010, Wernick2004}. By calibrating the detector with known radioactive sources, the energy of the incident gamma ray can be inferred from the spectral information derived from the electrical signal.

CsI:Tl and NaI:Tl are the most commonly used scintillation crystals for UAV-based gamma ray detection due to their low cost and good energy resolution \cite{Ardiny2024}. However, these crystals have limited sensitivity to neutrons. Neutron detection is important for identifying Special Nuclear Materials (SNM), such as enriched $^{235}$U or $^{239}$Pu, whose gamma emissions may be weak or shielded \cite{Marques2021}. NaIL (95\% $^6$Li enriched lithium co-doped NaI:Tl) is an emerging scintillation material capable of measuring gamma rays and neutrons with minimal processing electronics through pulse height analysis \cite{Yang2017}. This dual-mode detection capability enables SNM identification when gamma rays are heavily attenuated or below the low-level detection threshold.

A novel SiPM-based NaIL scintillation detection payload was developed and integrated with a multirotor UAV to enable aerial detection of gamma rays and neutrons. Section~\ref{sec:system_hardware} details the UAV-based radiation detection system integration. The system dead time and NaIL detector energy response are characterised in Section~\ref{sec:system_characterisation}. Section~\ref{sec:modelling_and_estimation} presents the derivation and experimental validation of a physically motivated radionuclide detection efficiency model, and describes how it can be applied to estimate ground-level radioactivity from aerial energy spectrum measurements. Section~\ref{sec:poc_flights} implements this method through a live radioactive source demonstration using the UAV-based radiation detection system. Finally, the discussion of the results and overall conclusion are presented in Sections~\ref{sec:discussion} and \ref{sec:conclusion} respectively.

\section{System Hardware}\label{sec:system_hardware}
\subsection{NaIL Scintillation Detector}\label{sec:nail_detector}
The SiPM-3000 from Bridgeport Instruments was selected for this study as it supports the use of custom scintillation crystals and features integrated read-out electronics, rugged detector housing, and a Broadcom AFBR-S4N66C013 SiPM array, which has a maximum photodetection efficiency of over 55\% \cite{Broadcom2023}. A 2-inch cylindrical NaIL scintillator from Luxium Solutions was optically bonded to the SiPM array in the SiPM-3000 using a 1 mm thick EJ-560 optical pad. It was positively pressure bonded to prevent decoupling of the scintillator from the SiPM array. To minimise noise and reduce the low-level detection threshold, the operating voltage (33 V), electronic gain (2), integration time (2 $\mu$s), dead time (3 $\mu$s), pulse trigger (7 mV), and noise trigger (3 mV) were set through the SiPM-3000 FGPA. The NaIL-SiPM-3000 detector, hereafter referred to as the NaIL scintillation detector, was calibrated by measuring the photopeak response from $^{133}$Ba, $^{137}$Cs, and $^{22}$Na sources over 5 minutes.

\subsection{UAV-based Radiation Detection System}\label{sec:detection_system}
\begin{figure}[h!]
    \centering
    \includegraphics[width=0.85\columnwidth]{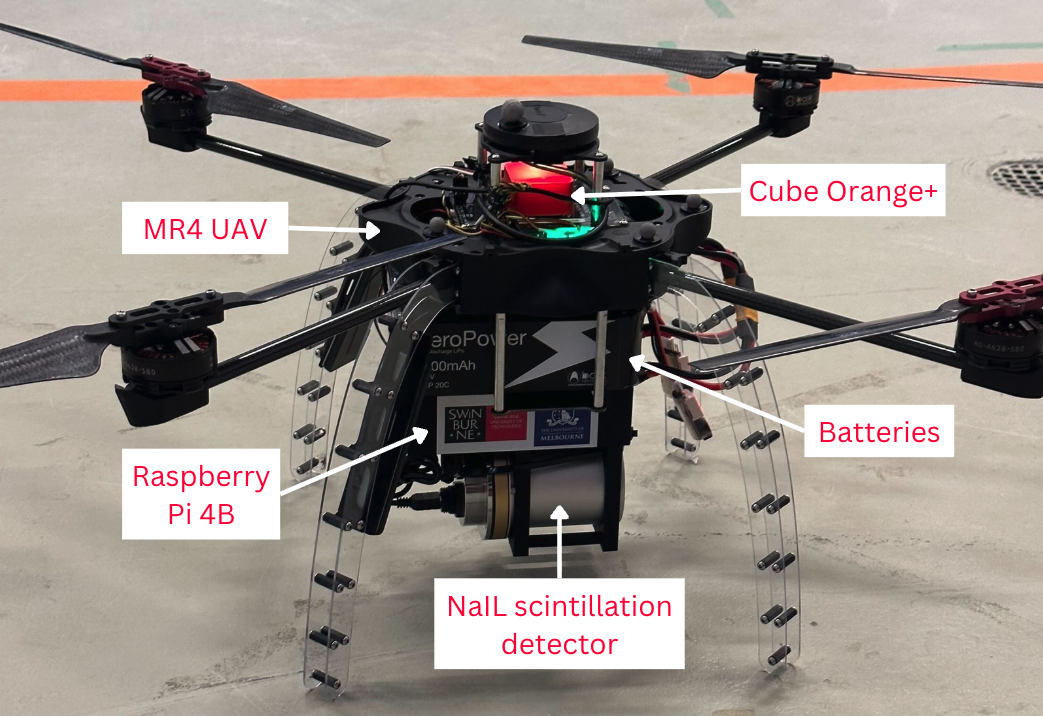}
    \caption{SiPM-based NaIL scintillation detection payload integrated with the MR4 quadrotor UAV from Bask Aerospace.}
    \label{fig:payload_with_drone}
\end{figure}

Aerial gamma ray and neutron detection was facilitated by mounting a SiPM-based NaIL scintillation detection payload on a MR4 quadrotor UAV from Bask Aerospace, as shown in Figure~\ref{fig:payload_with_drone}. The MR4 aerial platform featured two 8,000 mAh 14.8 V LiPo batteries, supporting flight for up to 15 minutes (with the 1 kg payload), and a Cube Orange+ flight controller running ArduPilot firmware. The payload consisted of four components: (1) the NaIL scintillation detector from Section~\ref{sec:nail_detector}, (2) a Raspberry Pi 4B, (3) a 5,000 mAh power bank, and (4) a 3D printed case containing components (1) to (3). The NaIL scintillation detector was operated via the Pi using the open-source Python-based wxMCA software \cite{SiPM3000}. The Pi enabled autonomous control by sending MAVLink\footnote{MAVLink is a communication protocol commonly used with unmanned vehicles \cite{MAVLink2025}.} commands to the flight controller via a custom Python script. To support controlled indoor operation without GPS, the system was flown in a motion capture laboratory equipped with high-precision cameras that tracked reflective markers on the MR4 platform. Platform position data was transmitted to the Pi via Wi-Fi, then relayed to the flight controller at 10 Hz over MAVLink, effectively serving as a GPS replacement.

\unskip

\begin{figure}[h!]
\centering
\includegraphics[width=0.99\linewidth]{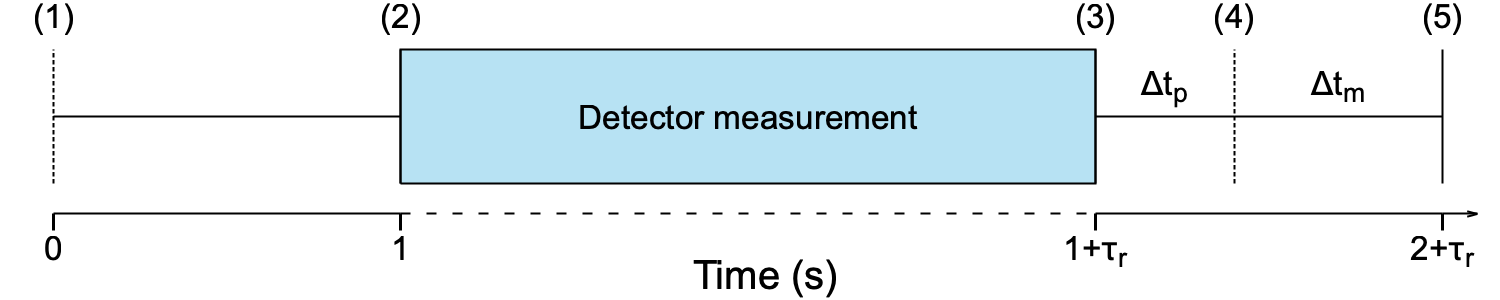}
\caption{NaIL scintillation detector measurement process. A new thread is created in step (1) and an initialisation message is sent to the MultiChannel Analyzer (MCA) in preparation for a new measurement. After a 1 s delay, allowing for the SiPM operating voltage to ramp up and stabilise, radiation is measured from step (2) to step (3) over the detector run time $\tau_r$. The detector data is retrieved by the thread after time delay $\Delta t_p$ in step (4), and saved by the main thread after $\Delta t_m$ in step (5). Step (1) is repeated after 1 s for the platform operating in the circular ArduPilot flight mode or until the system reaches the target waypoint in guided mode.}
\label{fig:measurement_process}
\end{figure}

The Python script on the Pi monitored the MAVLink heartbeat messages that were transmitted by the flight controller to detect whether the platform was operating in the guided or circular ArduPilot flight modes. If either mode was active, the script autonomously managed the NaIL scintillation detector and transmitted MAVLink waypoint commands to the flight controller, while still permitting manual piloting if needed. Figure~\ref{fig:measurement_process} outlines the NaIL scintillation detector measurement process within the Python script.

\unskip

\section{System Characterisation}\label{sec:system_characterisation}
\subsection{NaIL Scintillation Detector Energy Response}\label{sec:energy_response}
\begin{figure}[h!]
\includegraphics[width=0.495\linewidth]{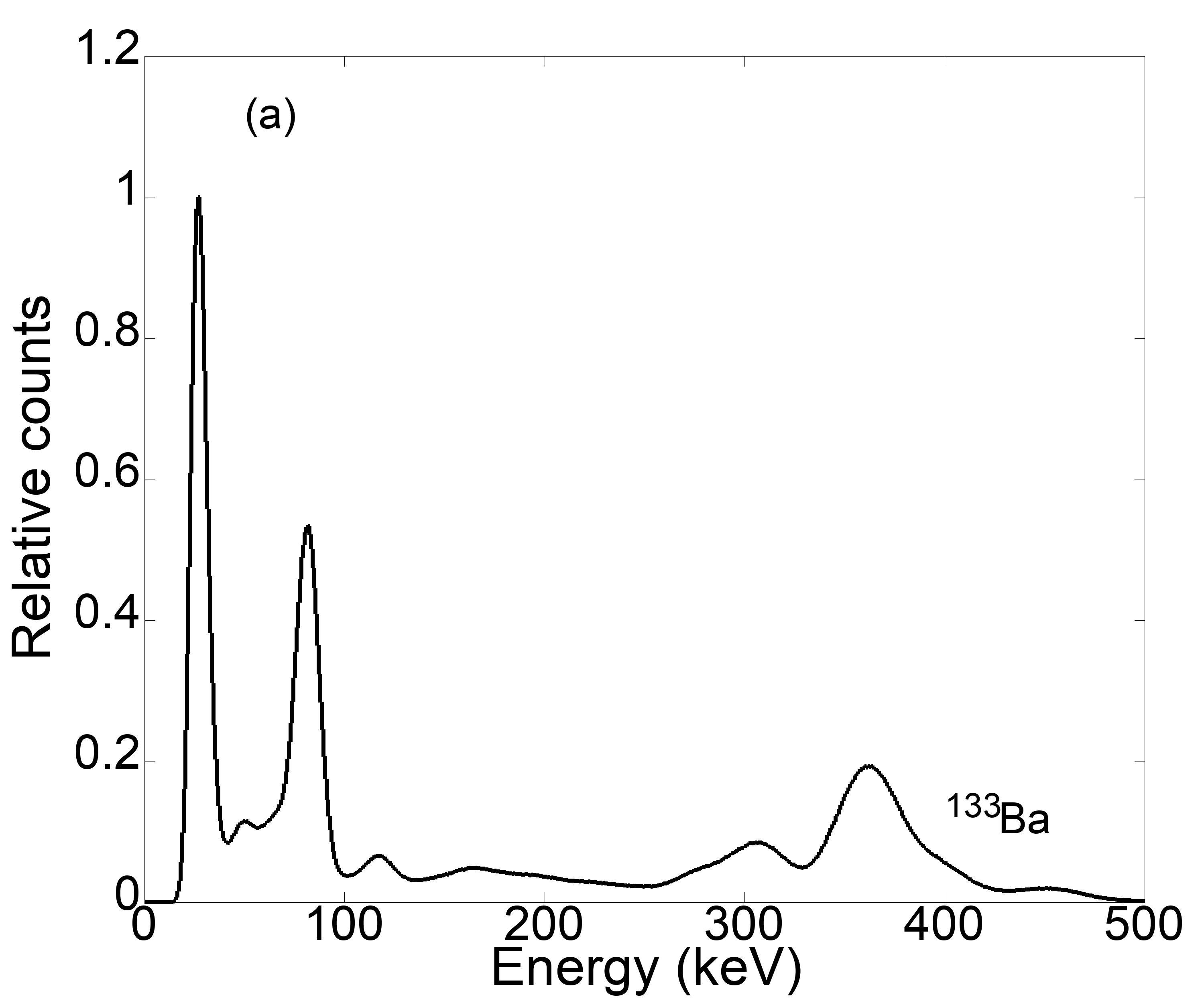}
\includegraphics[width=0.495\linewidth]{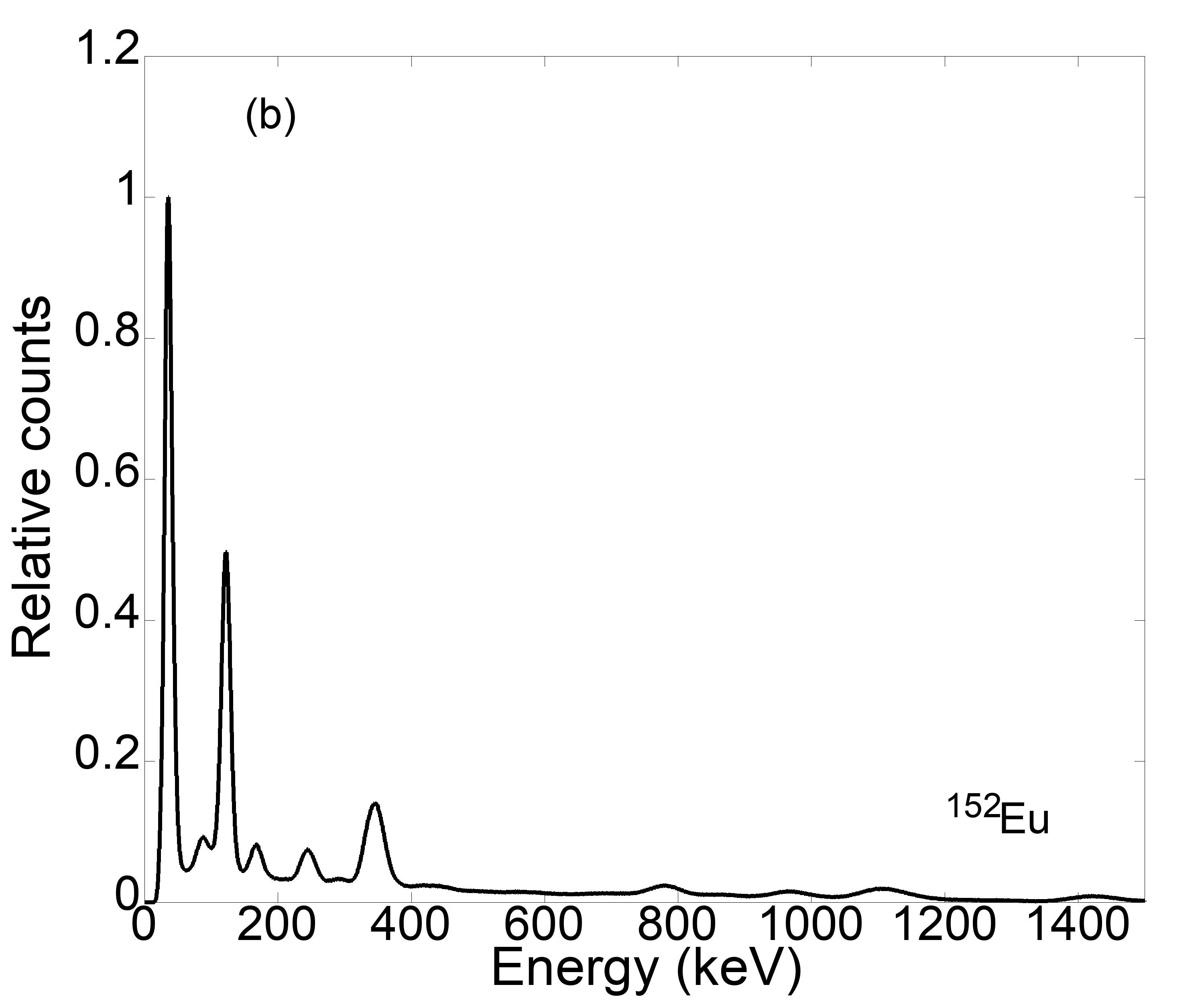}
\includegraphics[width=0.495\linewidth]{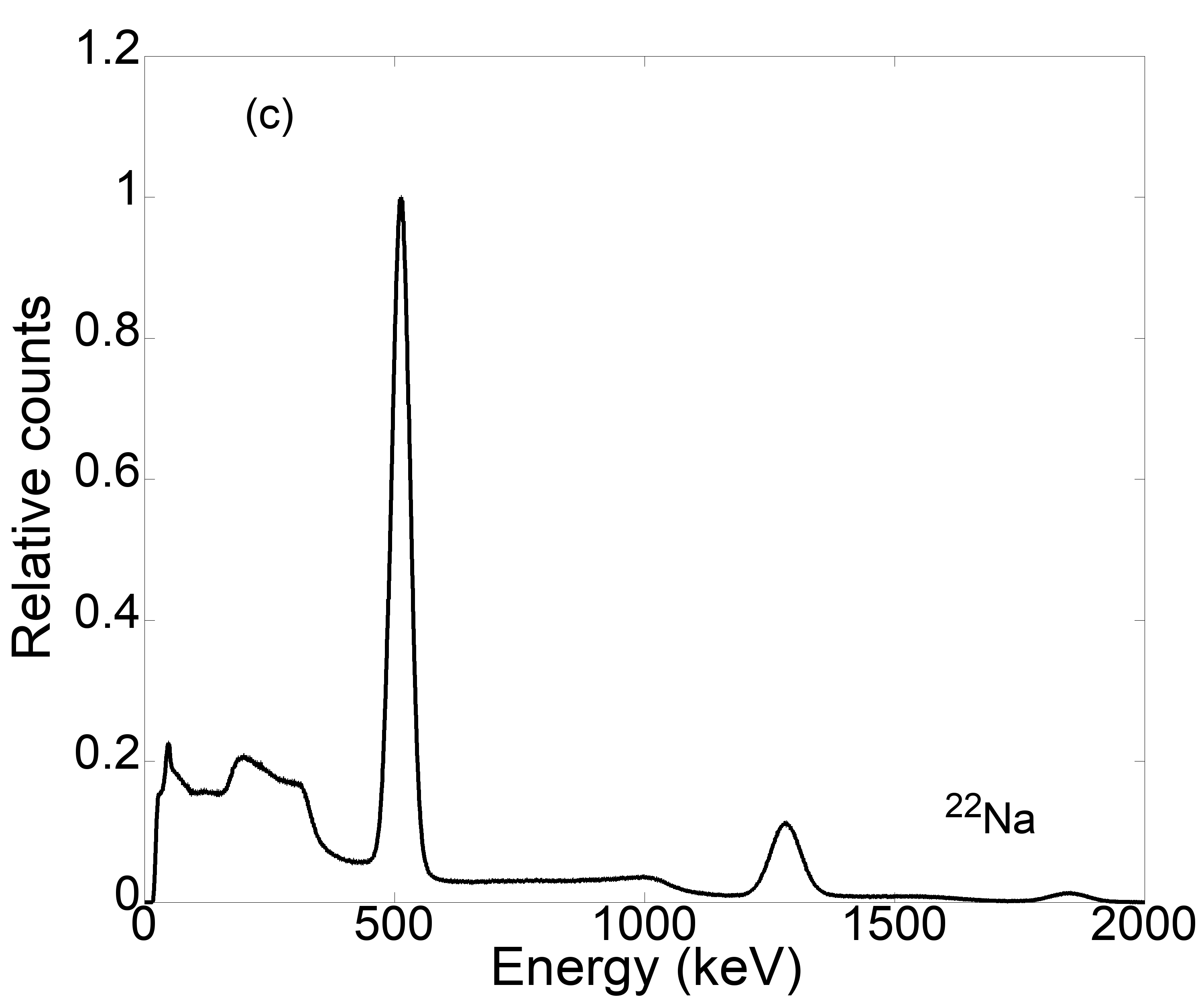}
\includegraphics[width=0.495\linewidth]{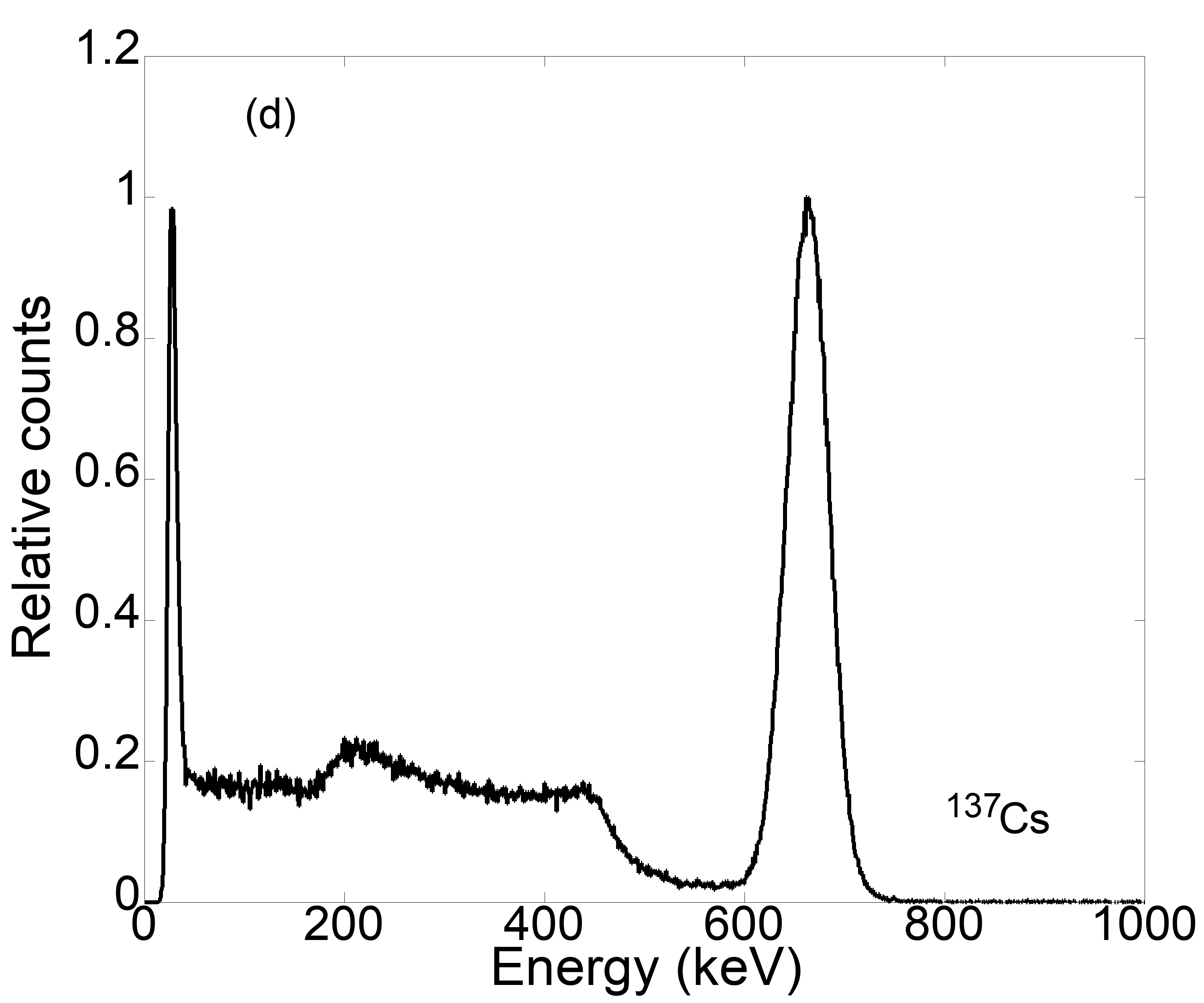}
\includegraphics[width=0.495\linewidth]{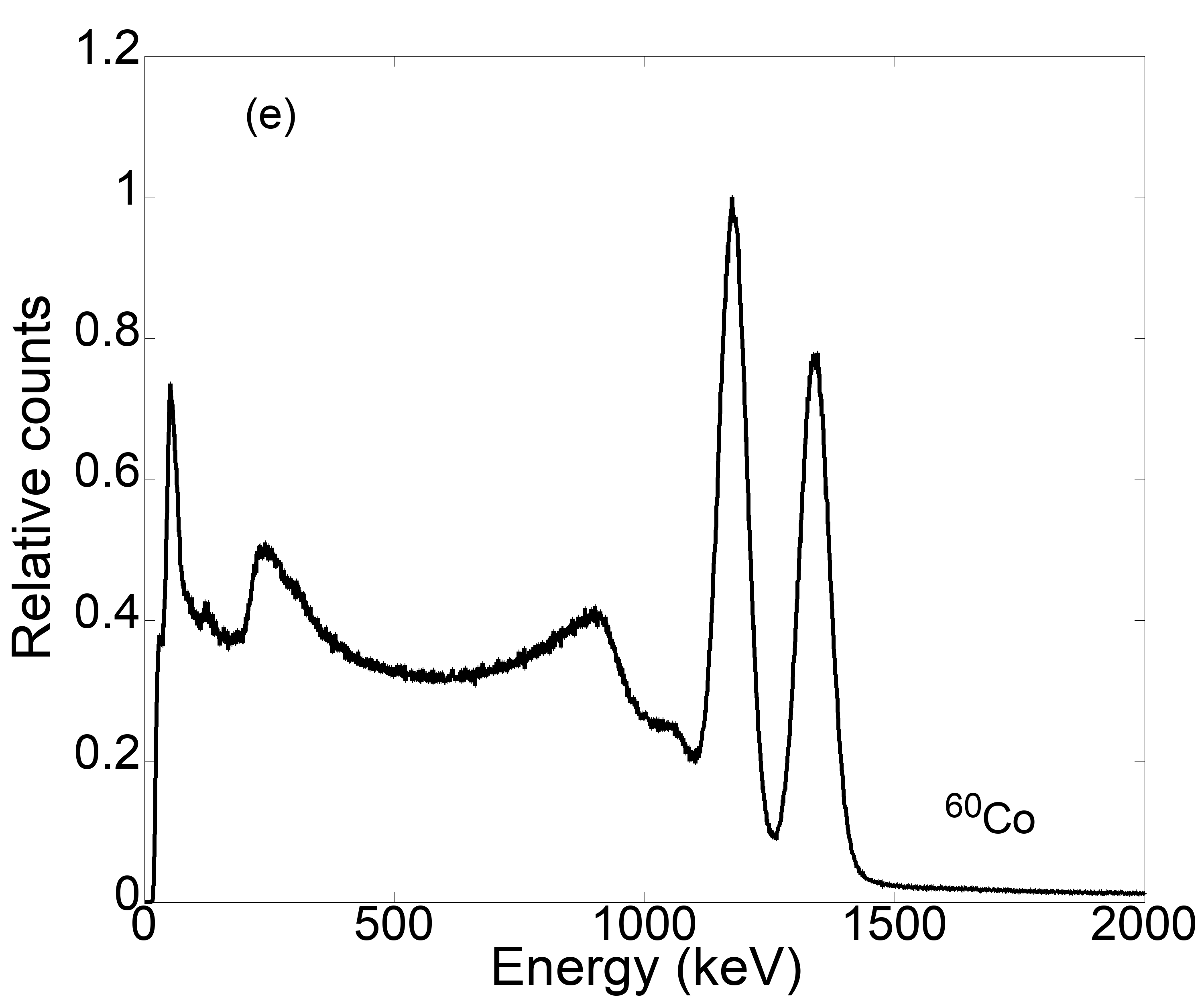}
\includegraphics[width=0.495\linewidth]{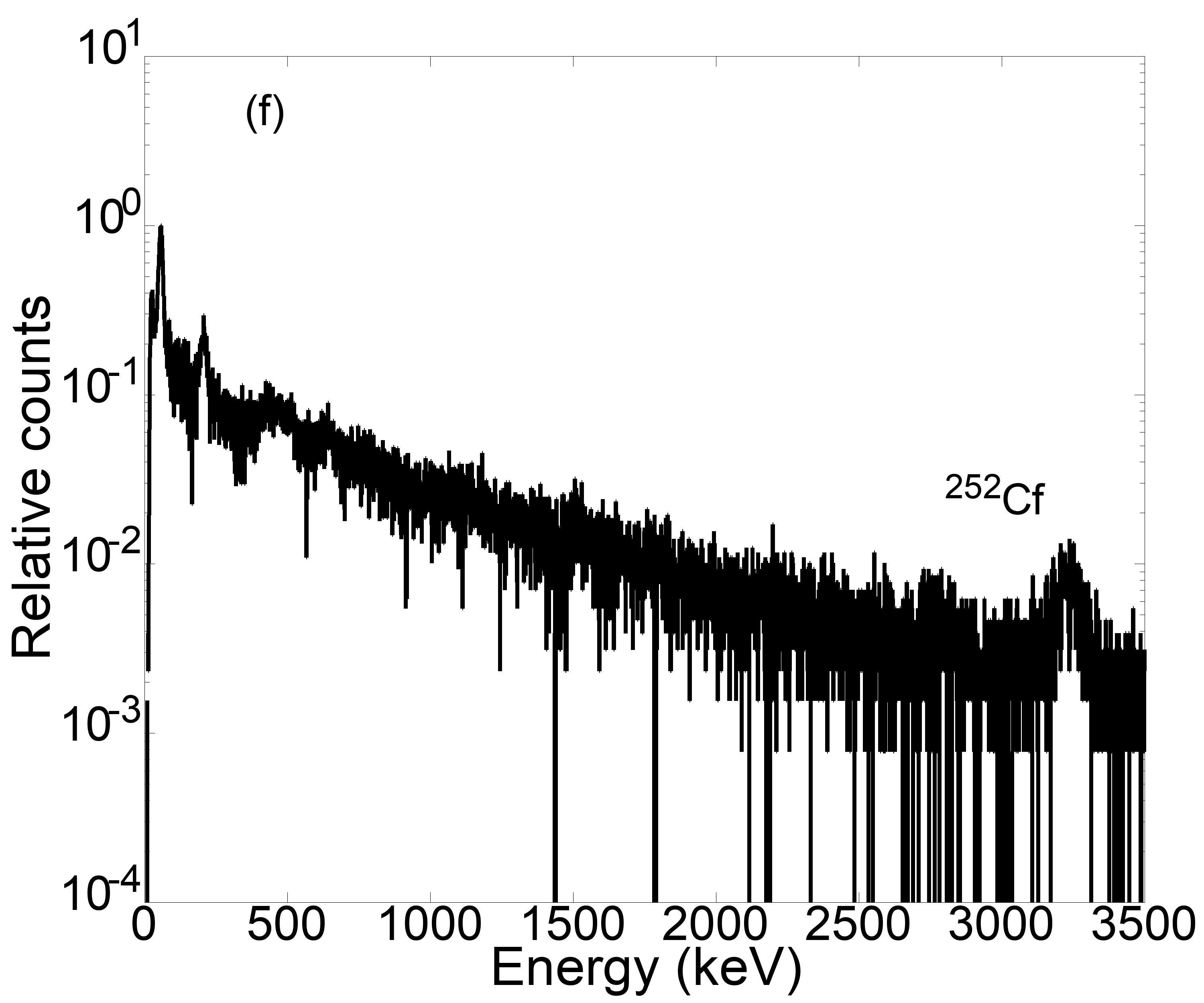}
\caption{Experimental $^{133}$Ba (\textbf{a}), $^{152}$Eu (\textbf{b}), $^{22}$Na (\textbf{c}), $^{137}$Cs (\textbf{d}), $^{60}$Co (\textbf{e}), and $^{252}$Cf (\textbf{f}) energy spectra measured with the NaIL scintillation detector.}
\label{fig:nail_spectra}
\end{figure}

\begin{figure}[h!]
    \centering
    \includegraphics[width=\columnwidth]{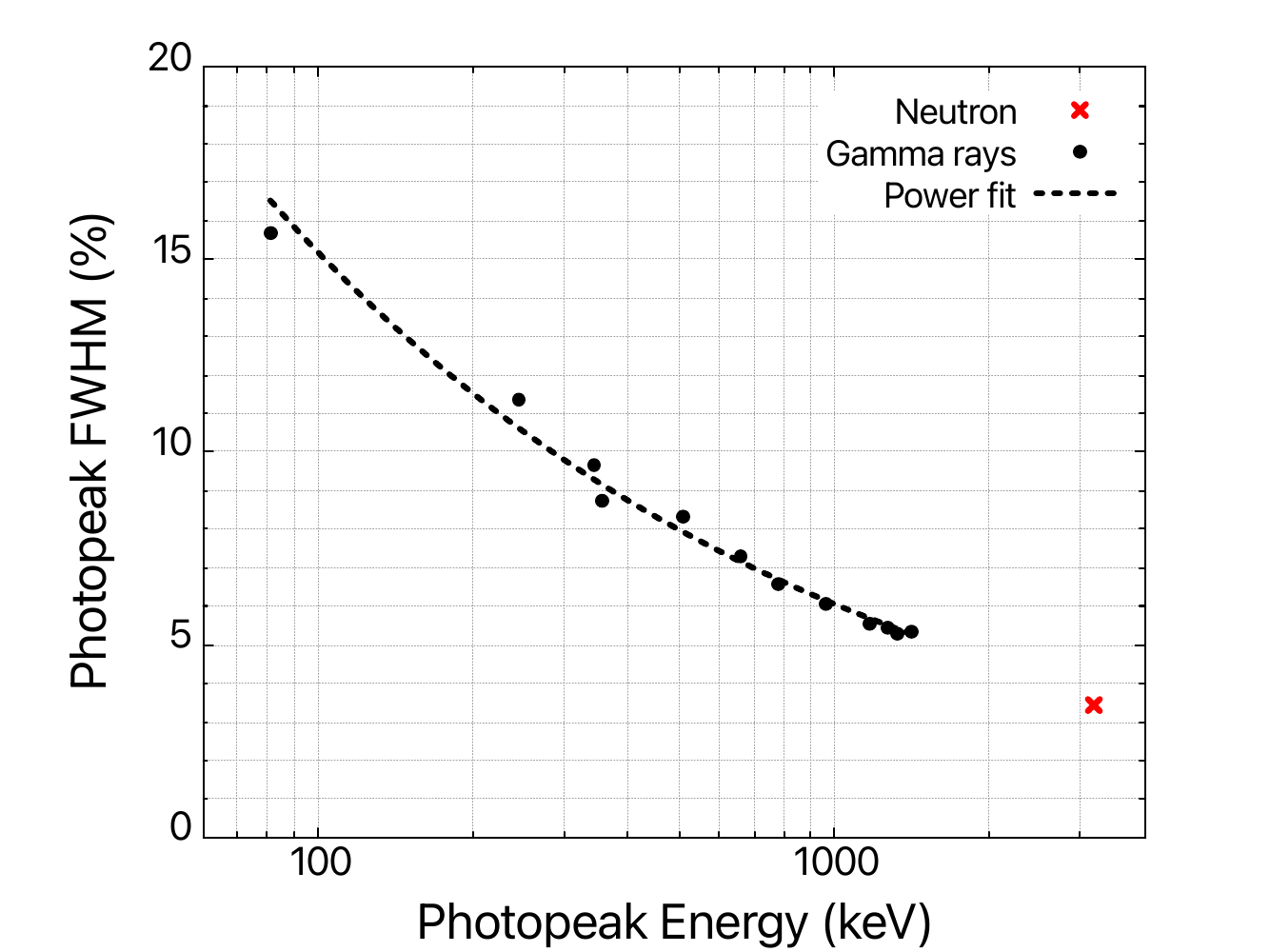}
    \caption{Photopeak Full Width Half Maximum (FWHM) derived from the energy spectra in Figure~\ref{fig:nail_spectra}. The points correspond to the FWHM for gamma ray energies between 81 keV ($^{133}$Ba) and 1408 keV ($^{152}$Eu), whereas the x-mark at 3200 keV is the full energy equivalent photopeak FWHM from $^6$Li neutron capture in the NaIL scintillator. The power function (dotted-line) fitted to the gamma ray photopeak FWHMs takes the form of FWHM = $(95 \pm 8) \cdot \gamma$-energy$^{(-0.40 \pm 0.01)}$.}
    \label{fig:energy_resolution}
\end{figure}

Dual-mode gamma ray and neutron detection was demonstrated by measuring the gamma ray and neutron emission from lab-based radioactive sources. The gamma ray sources were $^{133}$Ba, $^{152}$Eu, $^{22}$Na, $^{137}$Cs, and $^{60}$Co, whereas the gamma-neutron source was $^{252}$Cf. The corresponding energy spectra are shown in Figure~\ref{fig:nail_spectra}. Neutron capture with $^6$Li in the NaIL scintillator caused a full energy equivalent photopeak at 3200 keV in the $^{252}$Cf energy spectrum (Figure~\ref{fig:nail_spectra}f). The photopeak Full Width Half Maximum (FWHM) for the primary peaks in each energy spectrum is shown in Figure~\ref{fig:energy_resolution}. A power function was fitted to the photopeak FWHM values for the gamma ray emitting sources.

\subsection{System Dead Time}\label{sec:dead_time_characterisation}
\begin{figure}[h!]
\includegraphics[width=0.495\linewidth]{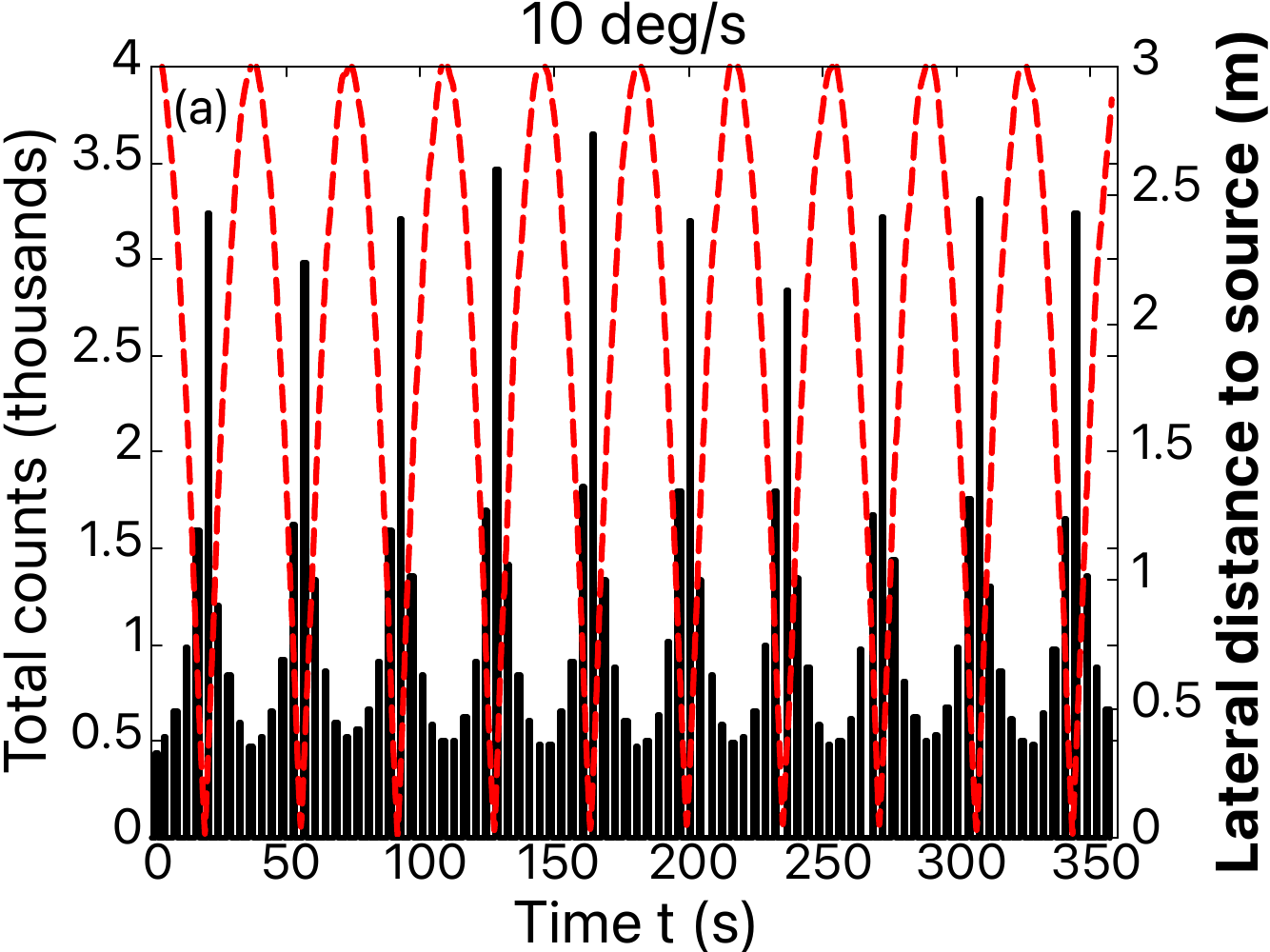}
\includegraphics[width=0.495\linewidth]{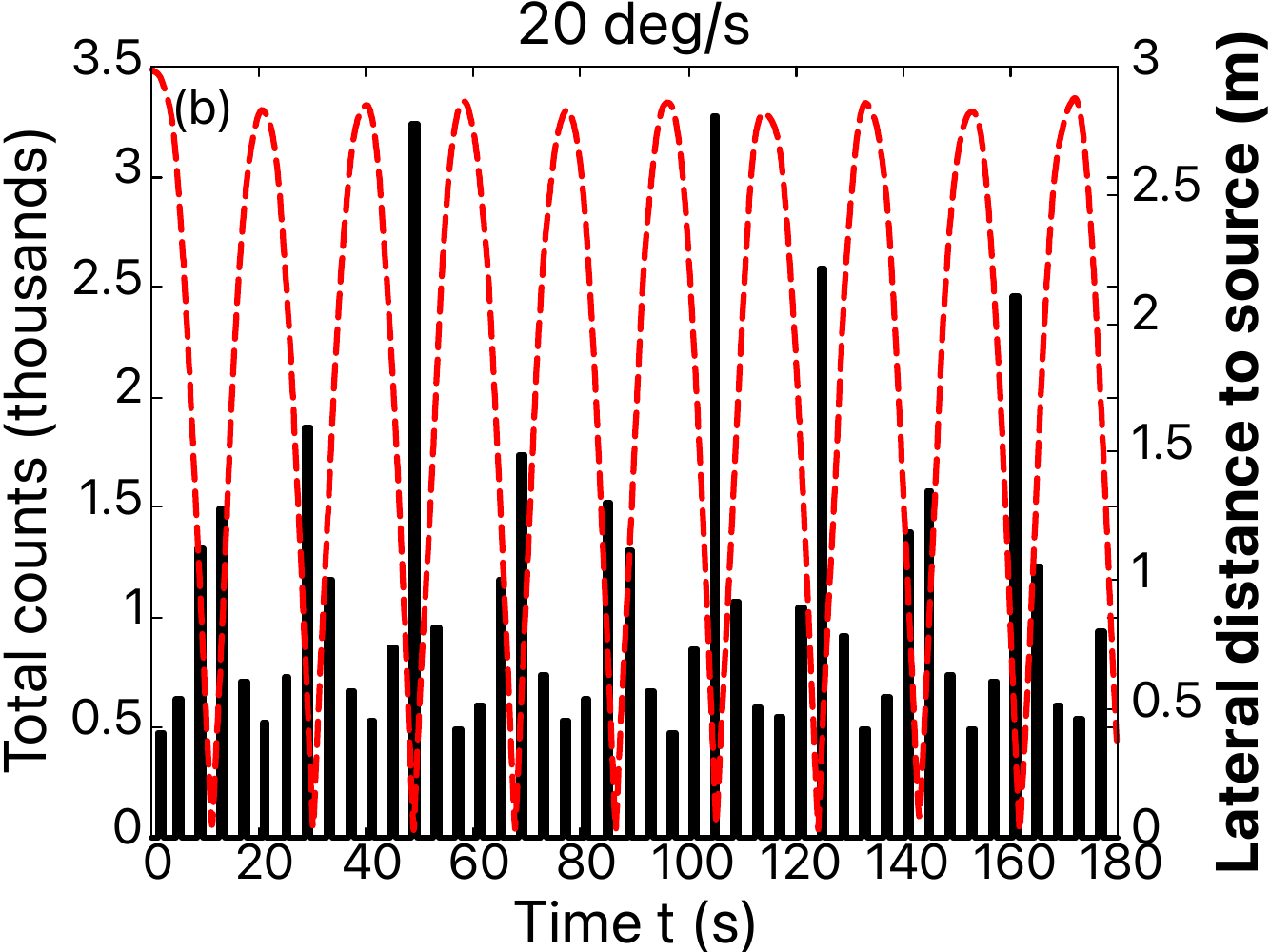}
\includegraphics[width=0.495\linewidth]{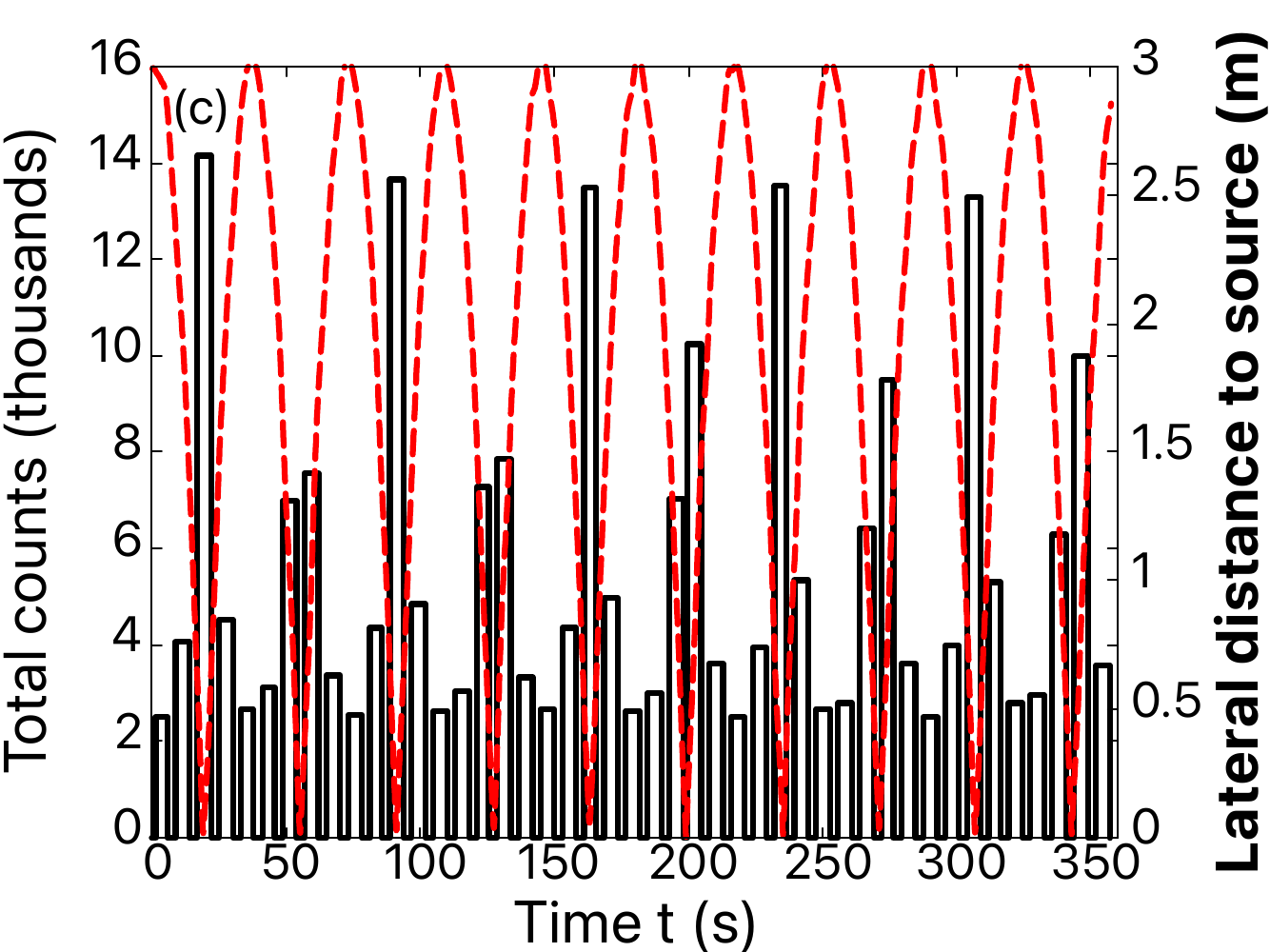}
\includegraphics[width=0.495\linewidth]{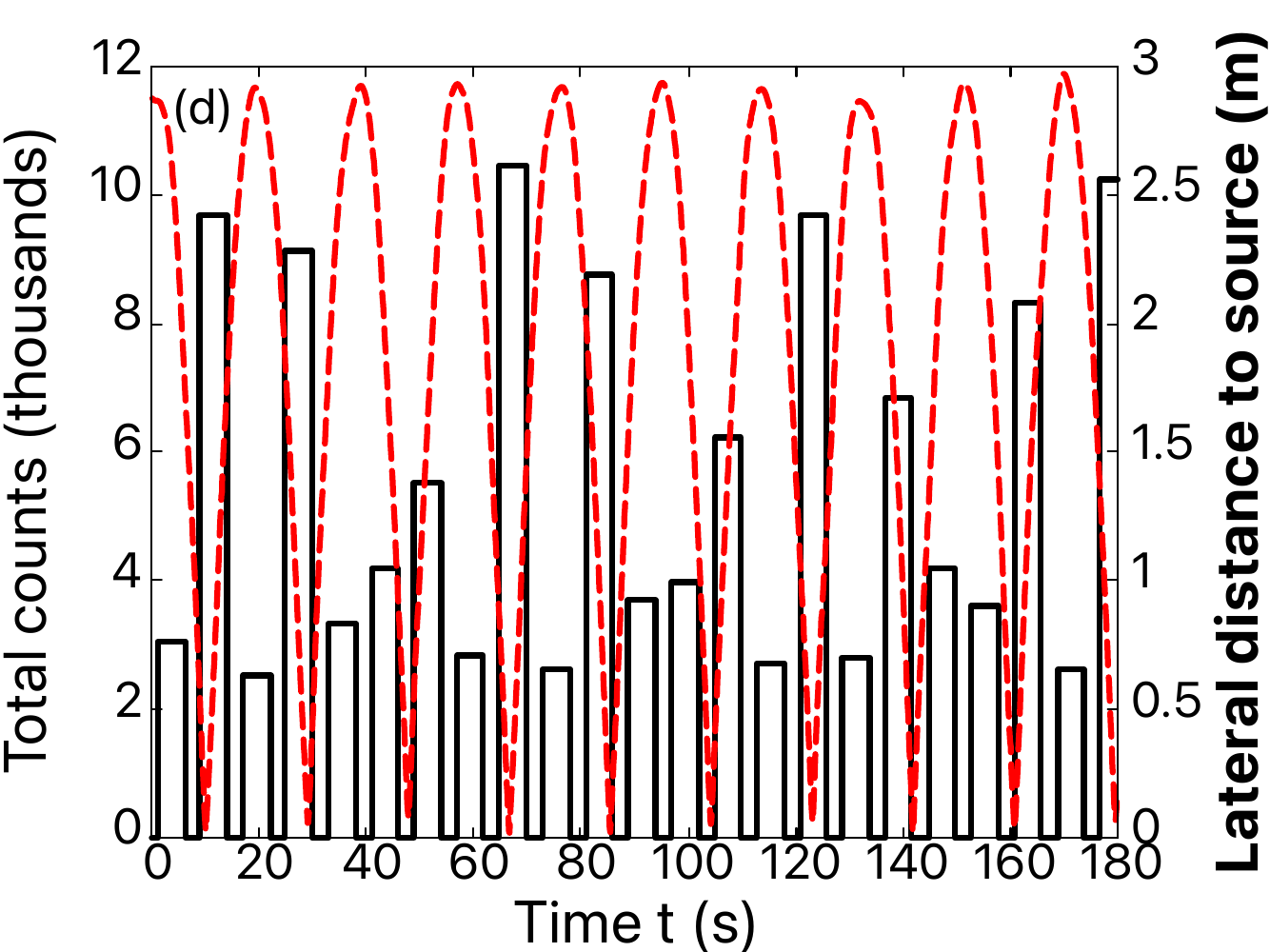}
\includegraphics[width=0.495\linewidth]{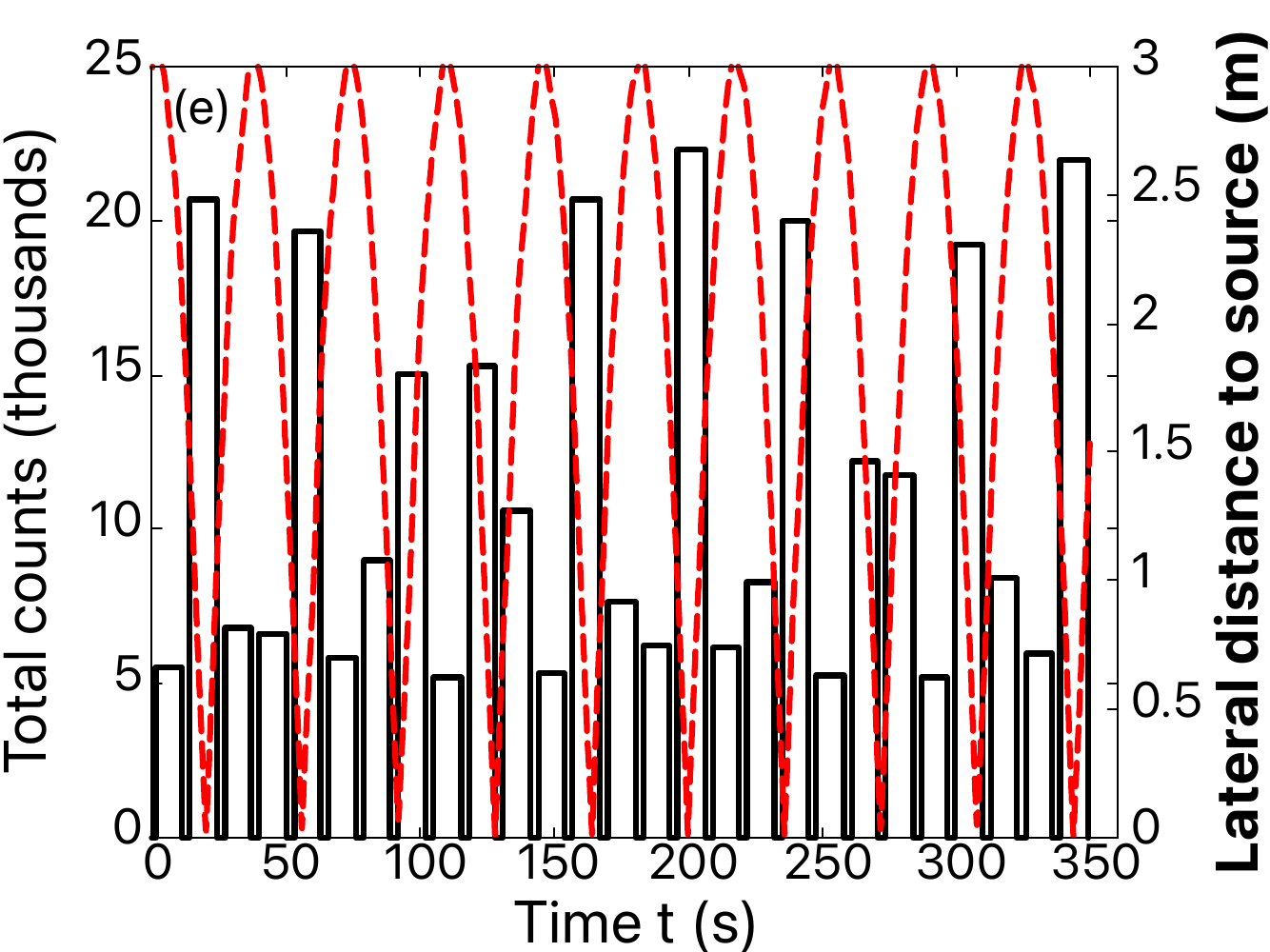}
\includegraphics[width=0.495\linewidth]{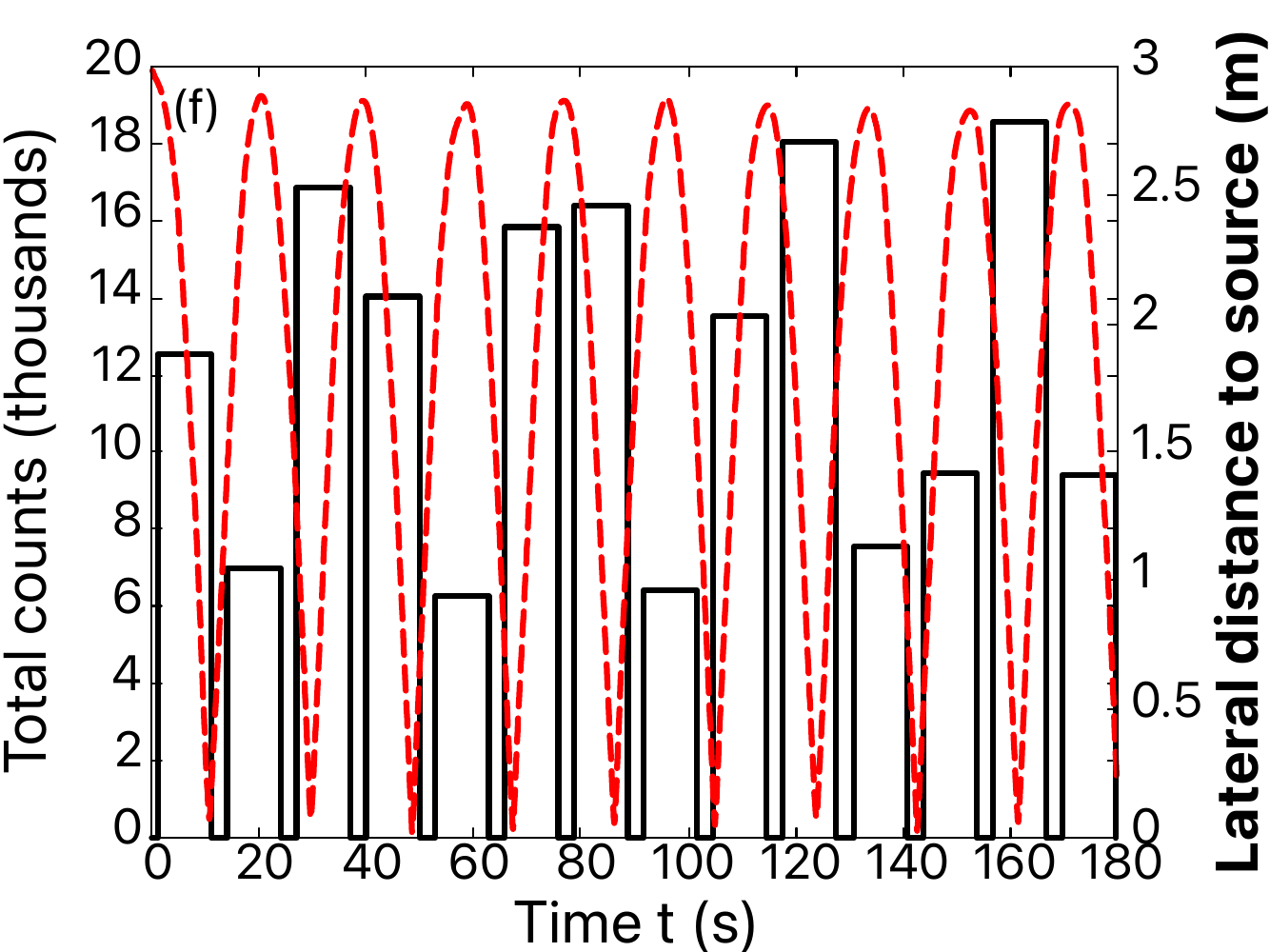}
\caption{Total count (step plot) and lateral distance from the radioactive source (dashed line) along the circular flight path at 2 m flight height. The detector measured radiation over 1 s (\textbf{a})-(\textbf{b}), 5 s (\textbf{c})-(\textbf{d}), and 10 s (\textbf{e})-(\textbf{f}) windows for flight speeds of 10 deg/s (left) and 20 deg/s (right) around a 3 m diameter circle.}
\label{fig:circular_cps_vs_time}
\end{figure}

System dead time refers to the total time the radiation detector is not actively measuring radiation from the surrounding environment. Characterising it is important because longer dead times reduce the amount of data collected, limiting the information gained about the environment. The system dead time consists of two components:
\begin{enumerate}
    \item \textit{Control Dead Time}: The interval during which the detector is deliberately disabled for configuration, data retrieval, and while the platform is moving to the target waypoint.
    \item \textit{Detector Dead Time}: The minimum time separation between incident gamma rays that allows them to be recorded as separate events.
\end{enumerate}
The dead time of the UAV-based radiation detection system was characterised through circular flights at a height of 2 m. During these flights, the system operated via the circular ArduPilot flight mode where it measured radiation along a 3 m diameter circle for 10 revolutions. A collimated $^{137}$Cs (97 MBq) source was placed on the ground under the circular trajectory. There were a total of six tests where the circular flight speeds were set to 10 deg/s and 20 deg/s with detector run times $\tau_r=$ 1 s, 5 s, and 10 s. Figure~\ref{fig:circular_cps_vs_time} displays the total counts (primary y-axis) and lateral distance from the radioactive source (secondary y-axis) in real time for each flight. The measured count spiked when the lateral distance was zero, indicating that the platform was above the collimated source. 

\unskip

The control dead time for the platform operating in the circular flight mode was $2.98 \pm 0.01$ s. As shown in Figure~\ref{fig:measurement_process}, the control dead time consisted of four sources: (1) the 1 s delay to allow the SiPM operating voltage to ramp up and stabilise, (2) the data retrieval delay $\Delta t_p$, (3) the data save delay $\Delta t_m$, and (4) the 1 s delay before the next measurement. The uncertainty in the control dead time was attributed to the variation in the MAVLink heartbeat message frequency. The detector dead time was between 0.1\% and 1\% of the detector run time, which was negligible in comparison to the control dead time.

\section{Modelling and Estimation}\label{sec:modelling_and_estimation}
\subsection{Radionuclide Detection Efficiency Model Derivation}\label{sec:detector_model}
\begin{figure}[h!]
    \centering
    \includegraphics[width=0.75\columnwidth]{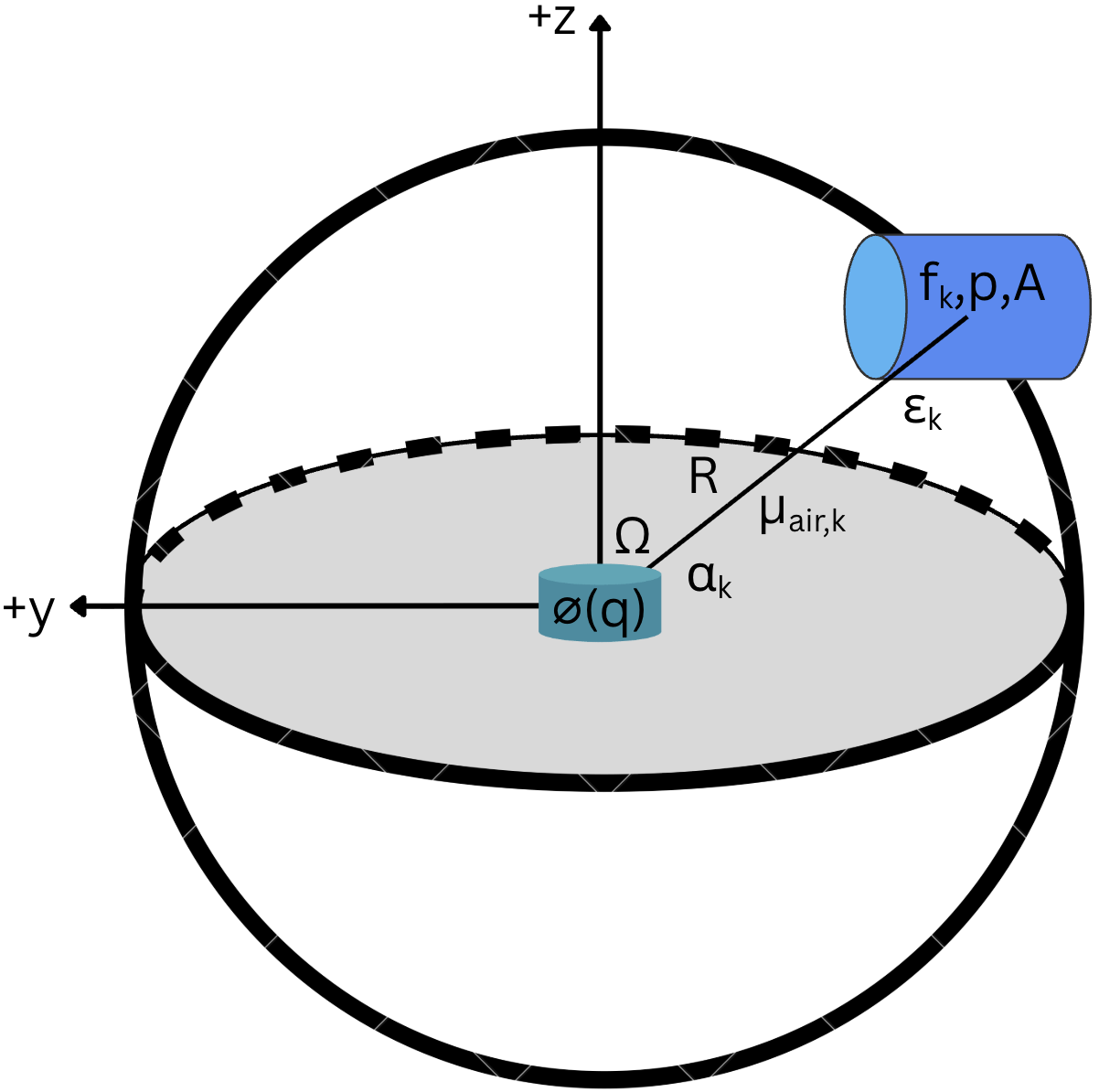}
    \caption{Sphere centered at the radioactive source location $q$ with activity $\phi(q)$. The radius $R$ denotes the distance to the detector at position $p$. A gamma ray of energy $E_k$ and branching ratio $\alpha_k$ is emitted within the solid angle $\Omega$ subtended by the effective surface area of the detector $A$. As the gamma ray propagates through air towards the detector, it is attenuated according to the absorption coefficient $\mu_{\text{air},k}$. The attenuated gamma ray contributes to the Counts Per Second (CPS) of photoelectrically absorbed gamma rays $f_k$ with a probability given by the intrinsic peak efficiency $\epsilon_k$.}\label{fig:interaction_probability_formulation}
\end{figure}

Let $q\in\mathbb{R}^3$ denote the position of a radioactive source with $N_k$ detectable characteristic gamma rays in a region of the ground plane located in the volume $Q\subseteq\mathbb{R}^3$, as illustrated in Figure~\ref{fig:interaction_probability_formulation}. The distribution of the source (in Bq) is denoted by $\phi\colon Q \rightarrow \mathbb{R}$. The probability that the $k$th characteristic gamma ray with energy $E_k$ is photoelectrically absorbed by a radiation detector at position $p\in\mathbb{R}^3$ may be expressed as the product of three distinct probabilities: (1) the radionuclide decays and produces a gamma ray with energy $E_k$, (2) the gamma ray reaches and enters the active volume of the detector, and (3) it is absorbed via the photoelectric effect within the scintillation crystal. These components are parameterised by the branching ratio $\alpha_k$, the detector solid angle $\Omega\colon \mathbb{R}\rightarrow\mathbb{R}$, which is a function of the distance between the detector and the radioactive source, the air absorption coefficient $\mu_{\text{air},k}$, and the intrinsic peak efficiency $\epsilon_k$. The Counts Per Second (CPS) of photoelectrically absorbed gamma rays with energy $E_k$ is modelled as 
\begin{equation}
    f_k(p,q) = \alpha_k\epsilon_k\phi(q)\Omega(||p-q||)\exp(-\mu_{\text{air},k}||p-q||),\label{eq:single_source}
\end{equation}
where the exponential term accounts for the air attenuation between the source and detector according to the Beer-Lambert law \cite{Swinehart1962}. The detector solid angle in the far field can be approximated as
\begin{equation}
    \Omega(R) = \frac{A}{4\pi R^2},\label{eq:solid_angle}
\end{equation}
where $A$ is the effective surface area of the detector from the perspective of the source and $R$ is the distance to its center of mass \cite{Knoll2010}. The effective area is treated as constant because the detector is assumed to be sufficiently high above the source.

\unskip

The intrinsic peak efficiency $\epsilon_k$ is proportional to the intrinsic efficiency $\epsilon_{\text{int} , k}$, which describes the probability of any interaction in the scintillation crystal. The relationship can be expressed as
\begin{equation}
    \epsilon_k = \epsilon_{\text{int}, k} \Big[\Big(\frac{\mu}{\rho}\Big)_{\text{pe}, k}+\kappa_k \Big(\frac{\mu}{\rho}\Big)_{\text{comp}, k}\Big]\Big(\frac{\mu}{\rho}\Big)_{k}^{-1},\label{eq:peak_absorption_in_crystal}
\end{equation}
where $(\mu / \rho)_{k}$, $(\mu / \rho)_{\text{comp}, k}$, and $(\mu / \rho)_{\text{pe}, k}$ are the total, Compton scattering, and photoelectric mass absorption coefficients respectively; pair production is negligible for the energy region of interest \cite{Vidmar2001,Moens1981}. The coefficient $\kappa_k$ represents the fraction of gamma rays that initially Compton scatter within the scintillation crystal before being photoelectrically absorbed at lower energies, contributing to the photopeak. According to the Beer-Lambert law, the intrinsic efficiency $\epsilon_{\text{int} , k}$ is approximated as one minus the transmittance
\begin{equation}
    \epsilon_{\text{int}, k} = 1 - \exp\Big(-\Big(\frac{\mu}{\rho}\Big)_{k}\rho l \Big),\label{eq:absorption_in_crystal}
\end{equation}
where $\rho$ is the density of the scintillator and $l$ is the mean chord length through it. By Cauchy's theorem, $l=2rh/(h+r)$ for a cylindrical crystal with radius $r$ and height $h$ \cite{Cauchy1908}. 

The average energy for single or multi-Compton scattered gamma rays with incident energy $E_k$ is 
\begin{equation}
    E^\prime_{k} = E_k[1-0.985\exp(-425 \text{ keV}/E_k)],
\end{equation}
with corresponding mass absorption coefficients $(\mu / \rho)^\prime_{k}$, $(\mu / \rho)^\prime_{\text{comp}, k}$, and $(\mu / \rho)^\prime_{\text{pe}, k}$. The intrinsic efficiency for internally Compton scattered gamma rays, assuming a mean path length of $l/2$, is given by 
\begin{equation}
    \epsilon^\prime_{\text{int}, k} = 1 - \exp\Big(-\frac{1}{2}\Big(\frac{\mu}{\rho}\Big)^\prime_{k}\rho l \Big),
\end{equation}
which is used to calculate the proportion of Compton scattered gamma rays that contribute to the photopeak
\begin{equation}
    \kappa_k = \frac{\big(\frac{\mu}{\rho}\big)'_{\text{pe}, k}\epsilon_{\text{int}, k}'}{\big(\frac{\mu}{\rho}\big)'_{k}-\big(\frac{\mu}{\rho}\big)'_{\text{comp}, k}\epsilon_{\text{int}, k}'}.\label{eq:kappa_k}
\end{equation}
according to \"{O}zmutlu and Ortaovali, 1976 \cite{Ozmutlu1976}. Substituting this expression into Equation~\ref{eq:peak_absorption_in_crystal} enables the complete evaluation of the intrinsic peak efficiency $\epsilon_k$ using values that are available in standard reference tables.

The CPS of photoelectrically absorbed gamma rays emitted by a known radionuclide is obtained by summing the individual contributions from the characteristic gamma rays represented by Equation~\ref{eq:single_source} for all $k \in [1, 2, ..., N_k]$
\begin{equation}
    f(p, q) = \sum_{k=1}^{N_k}f_k(p,q)= \frac{A\zeta(||p-q||)\phi(q)}{||p-q||^2} ,\label{eq:multiple_source}
\end{equation}
where $\zeta(R) = \sum_{k=1}^{N_k}\alpha_k\epsilon_k\exp(-\mu_{\text{air},k}R)/(4\pi)$ for a specific radionuclide. In practice, the source position can not be isolated in a single volume detector measurement. The \textit{radionuclide count} is defined as the accumulation of all photoelectrically absorbed gamma rays emitted by a radionuclide within the volume $Q$ over the effective measurement time $\tau$, which is the detector run time $\tau_r$ corrected for the detector dead time. The measurement time is assumed to be much shorter than the half-life of the radioactive source. During the measurement period, the platform follows the trajectory $p(t)$ for $t \in [T-\tau, T]$, where $T$ denotes the wall-clock time that includes dead time. This leads to the general equation for the radionuclide count 
\begin{equation}
    F(p(\cdot),T,\tau) = \int_{T-\tau}^{T}\Bigg[\int_Q f(p(t),q) \ dq+B(p(t))\Bigg] \ dt,\label{eq:aggregate_source}
\end{equation}
where $B(p(t))$ is the CPS of background radiation along platform trajectory $p(t)$. If the background radiation along $p(t)$ is unknown, it can be incorporated into the source distribution $\phi(q)$ by setting the known background $B(p(t))$ to zero for all $p(t)$. Equation~\ref{eq:aggregate_source} can be simplified assuming a spatially-static measurement where $p(t) = \bar{p}$ for all $t \in [T-\tau, T]$
\begin{equation}
    y(\bar{p},\tau) = \tau\Bigg[\int_Q f(\bar{p},q) \ dq+B(\bar{p})\Bigg].\label{eq:static_aggregate_source}
\end{equation}

\subsection{Radionuclide Detection Efficiency Model Validation}\label{sec:radionuclide_detection_validation}
\begin{figure}[h!]
    \centering
    \includegraphics[width=\columnwidth]{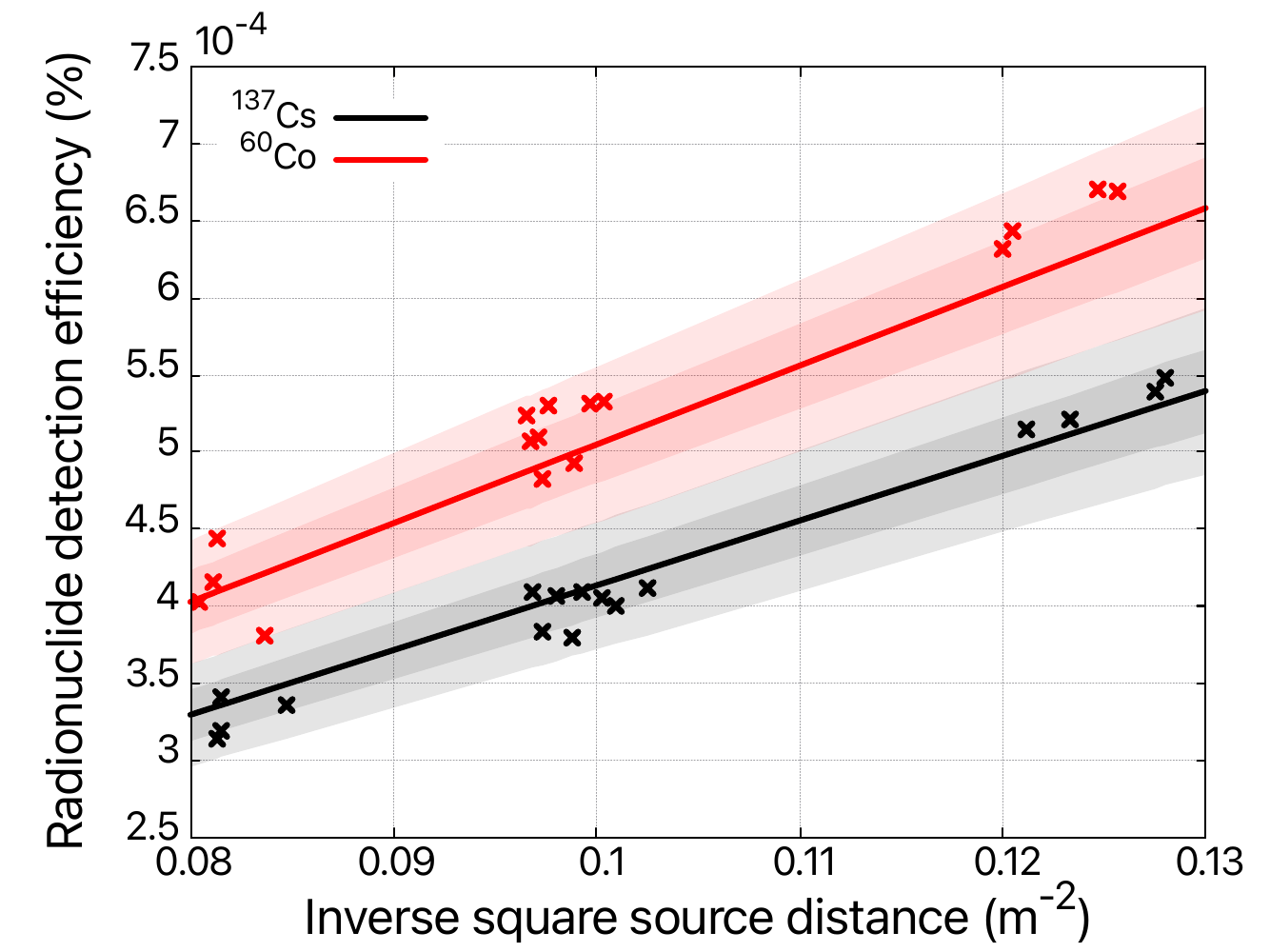}
    \caption{Source distance-dependent experimental (x-mark) and modelled (solid line) radionuclide detection efficiencies for $^{137}$Cs and $^{60}$Co gamma ray emissions. The inner and outer shaded bands around each solid line represent model uncertainties of 5\% and 10\% respectively.}
    \label{fig:model_verification}
\end{figure}

Experimental validation of the spatially-static radionuclide count model in Equation~\ref{eq:static_aggregate_source} was performed by individually measuring the spectral response to $^{137}$Cs (97 MBq) and $^{60}$Co (49 MBq) source emissions. Each source was positioned at the origin and detector measurements were taken over $\tau_r=37.5$ s at a height of 3 m on a 1 m-spaced 3 m $\times$ 3 m grid. The background radiation was subtracted from each measurement in the energy domain and a single or double Gaussian function (single for $^{137}$Cs and double for $^{60}$Co) with a quadratic background was fit to each photopeak using linear least squares. The radionuclide count was calculated from the counts within $\pm3$ standard deviations of the fitted Gaussian centroid(s). This count was then normalised by the product of the effective measurement time (from the SiPM-3000 output) and the source activity to calculate the radionuclide detection efficiency. Figure~\ref{fig:model_verification} presents a comparison of the modelled and measured efficiencies for the $^{137}$Cs and $^{60}$Co sources.

Using a scenario that replicated the experimental configuration, the modelled radionuclide detection efficiencies were calculated by normalising the radionuclide count rate (Equation~\ref{eq:static_aggregate_source} with $\tau=1$ s) by the source activity. In the model, a cylindrical detector with radius $r=2.54$ cm and height $h=5.08$ cm was used to match the dimensions of the physical NaIL scintillator. The effective surface area was approximated as a rectangle with area $A=2rh=25.8$ cm$^2$ as the detector was orientated along the y-axis to maximise the surface area from the perspective of the ground plane. The branching ratios and parameters used in the intrinsic peak efficiency $\epsilon_k$ (Equation~\ref{eq:peak_absorption_in_crystal}) are summarised in Table~\ref{table:branching_mass_absorption_params}. The branching ratios were obtained using the NuDat 3.0 database \cite{NNDC2025} and the mass absorption coefficients were from the NIST XCOM database using NaIL (assumed to be composed of 99.2\% NaI, 0.1\% Tl, and 0.7\% Li) \cite{NISTXCOM}. As shown in Figure~\ref{fig:model_verification}, all experimental radionuclide detection efficiencies were within 10\% of the model, as indicated by the outer shaded band.

\vspace{-3pt}
\begin{table}[h!]
\caption{Gamma ray energy $E_k$ dependent values for the branching ratio $\alpha_k$ \cite{NNDC2025}, total $(\mu / \rho)_k$, Compton scattering $(\mu / \rho)_{\text{comp}, k}$, and photoelectric $(\mu / \rho)_{\text{pe}, k}$ mass absorption coefficients \cite{NISTXCOM}, and the proportion of Compton events that contribute to the full energy peak $\kappa_k$. NaIL has a density of $\rho=3.66$ g/cm$^3$ \cite{Luxium2023}.}\label{table:branching_mass_absorption_params}
\centering
\setlength{\tabcolsep}{3.5pt}
\begin{tabular}{|c|c||c|c|c|c|c|}
\hline
Isotope & \makecell{$E_k$ \\ keV} & \makecell{$\alpha_k$ \\ \%} & \makecell{$(\mu/\rho)_k$ \\ $10^{-2}$ cm$^2$/g} & \makecell{$(\mu/\rho)_{\text{comp}, k}$ \\ $10^{-2}$ cm$^2$/g} & \makecell{$(\mu/\rho)_{\text{pe}, k}$ \\ $10^{-3}$ cm$^2$/g} & $\kappa_k$ \\
\hline
$^{137}$Cs & 662 & 85.10 & 7.658 & 6.541 & 8.518 & 0.34\\
\hline
\multirow{2}{*}{$^{60}$Co} & 1173 & 99.85 & 5.345 & 5.002 & 2.510 & 0.24\\
\cline{2-7}
& 1332 & 99.98 & 4.988 & 4.691 & 1.987 & 0.23 \\
\hline
\end{tabular}
\end{table}

\vspace{-3pt}
\subsection{Ground-Level Radioactivity Estimation}\label{sec:source_distribution_estimation}
The spatially-static radionuclide count model in Equation~\ref{eq:static_aggregate_source} can be generalised to $N$ measurements, indexed by $i \in \{1, 2, ..., N\}$. Each measurement is characterised by a sample time $T_i$, a measurement duration $\tau_i$, and an average measurement position $p(t_i)=\bar{p}_i$ for $t_i\in[T_i-\tau_i, T_i]$. The ground-level radionuclide activity, known as the source distribution $\phi(q)$, can be represented using different basis sets that are tailored to the requirements of the application. This work considers a grid of $M$ Dirac delta functions as the basis
\begin{equation}
\phi(q)=\sum_{j=1}^M\beta_j\delta(q-\mu_j),\label{eq:dirac_decomposition_phi}
\end{equation}
where the $j$th Dirac delta function is centered at $\mu_j\in Q$ within the ground plane and scaled with unknown magnitude $\beta_j\in\mathbb{R}_{\geq 0}$. The radionuclide counts $\textbf{y}=(y(\bar{p}_1, \tau_1), y(\bar{p}_2, \tau_2), ..., y(\bar{p}_N, \tau_N))^T$ from Equation~\ref{eq:static_aggregate_source} can be written in compact matrix form as
\begin{equation}
    \textbf{y}= H\boldsymbol{\beta}+\textbf{b},\label{eq:equation_to_solve}
\end{equation}
where $\textbf{b}=(B(\bar{p}_1)\tau_1, B(\bar{p}_2)\tau_2, ..., B(\bar{p}_N)\tau_N)^T$ represents the background measurements, $\boldsymbol{\beta} = (\beta_1, \beta_2, ..., \beta_M)^T$, and $H$ is an $N \times M$ matrix with $ij$th matrix element
\begin{equation}
    H_{ij} = \frac{A\tau_i\zeta(||\bar{p}_i-\mu_j||)}{||\bar{p}_i-\mu_j||^2}. \label{eq:source_activity_estimation_algorithm}
\end{equation}
The source distribution $\phi(q)$ parametrised by $\boldsymbol{\beta}$ in Equation~\ref{eq:dirac_decomposition_phi} can be estimated by solving the non-negative least squares optimisation problem
\begin{equation}
    \hat{\boldsymbol{\beta}} = {\arg \min}_{\boldsymbol{\beta} \geq \boldsymbol{0}} || H\boldsymbol{\beta} - \tilde{\textbf{y}}_{\text{c}} ||^2,\label{eq:nnls_optimisation}
\end{equation}
where $\tilde{\textbf{y}}_c$ is the background-corrected radionuclide count measurements.

\section{Flight Trials}\label{sec:poc_flights}
The functionality of the developed UAV-based radiation detection system and ground-level radioactivity estimation method introduced in Section~\ref{sec:source_distribution_estimation} was demonstrated through a controlled indoor flight trial supported by the Defence Science and Technology Group (Australia). There were two parts to this flight trial:
\begin{enumerate}
    \item \textit{Background Characterisation}: Preliminary measurements on a 1 m-spaced 3 m $\times$ 3 m grid showed constant background radiation at flight heights of 1.5 m, 3 m, and 4.5 m. The height dependence on background radiation was characterised through 2-minute measurements at these flight heights.
    \item \textit{Live Radioactive Source Demonstration}: Six flights were conducted over a 3 m x 3 m area using a 0.5 m-spaced raster scan at heights of 1.5 m, 3 m, and 4.5 m. A $^{137}$Cs (97 MBq) or $^{60}$Co (49 MBq) source was placed on the ground at $x=1.5$ m and $y=1.5$ m. Radiation was measured in discrete intervals of $\tau_r=12.24$ s over a total measurement duration of 10 minutes per flight.
\end{enumerate}
For both parts of the trial, the system was launched manually in the position hold ArduPilot flight mode. The pilot transitioned to the guided flight mode once the platform stabilised in the air, triggering the automated waypoint navigation sequence managed by the onboard Python script. The platform hovered within 0.1 m of the target waypoint during data collection and proceeded to the next waypoint upon completion of the measurement.

\begin{figure}[h!]
\includegraphics[width=0.495\linewidth]{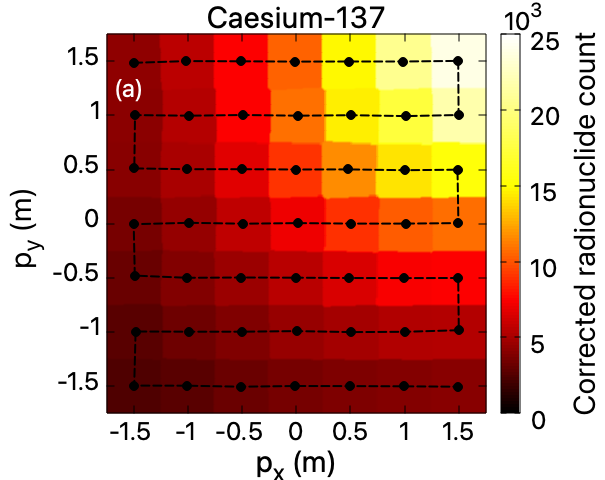}
\includegraphics[width=0.495\linewidth]{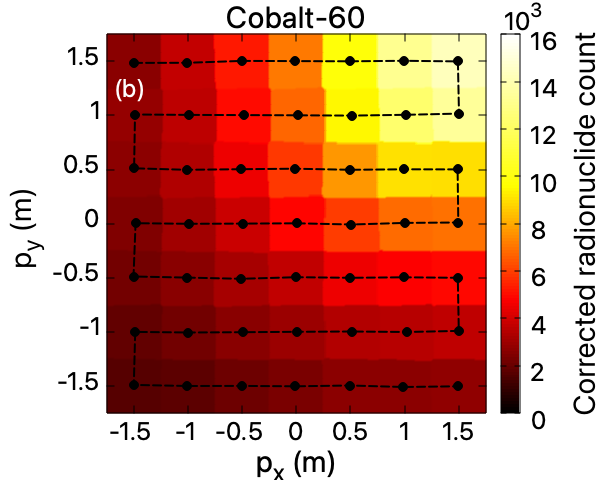}
\includegraphics[width=0.495\linewidth]{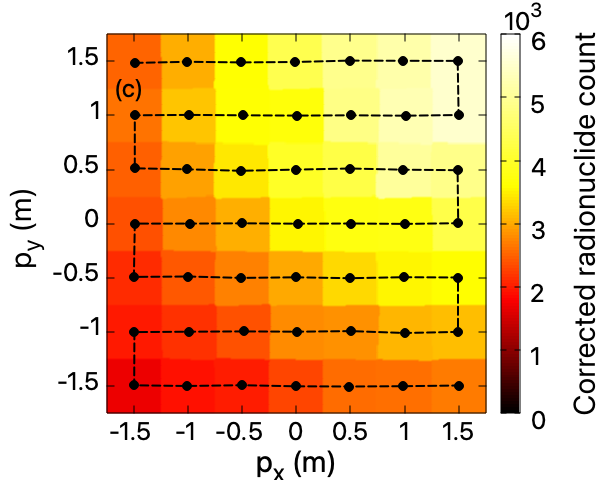}
\includegraphics[width=0.495\linewidth]{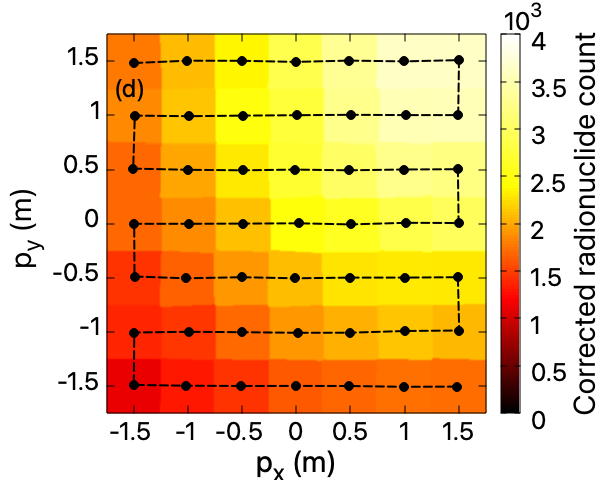}
\includegraphics[width=0.495\linewidth]{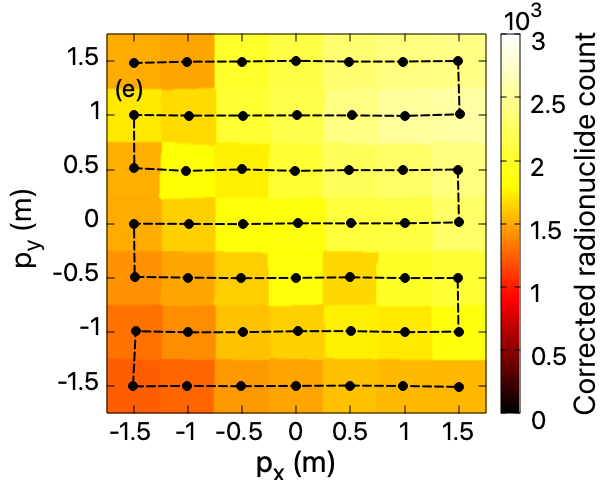}
\includegraphics[width=0.495\linewidth]{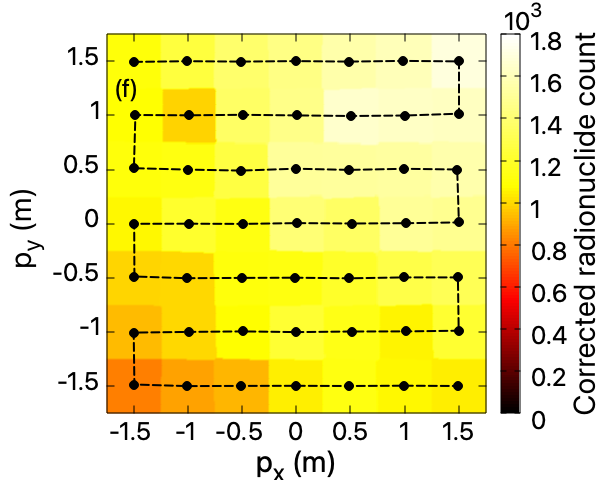}
\caption{Background-corrected radionuclide count heatmaps from the live radioactive source demonstration. Measurements (dotted-dashed line) were obtained at discrete points along the raster trajectory for flight heights of 1.5 m (\textbf{a})-(\textbf{b}), 3 m (\textbf{c})-(\textbf{d}), and 4.5 m (\textbf{e})-(\textbf{f}). The figures in the left column correspond to the $^{137}$Cs (97 MBq) flights, whereas the right column represents the $^{60}$Co (49 MBq) flights.}
\label{fig:at_altitude_radiation_map}
\end{figure}

Background-corrected energy spectra were obtained by subtracting background radiation, scaled to the live source duration ($\tau_i = 12.16 \pm 0.01$ s for all $i\in\{1, 2, ..., N\}$), from the spectra measured during the live radioactive source demonstration. Photopeaks within the background-corrected energy spectra were fitted using a quadratic background combined with a single Gaussian function for $^{137}$Cs or a double Gaussian function for $^{60}$Co. The fit was performed within an energy window of $\pm 5$ standard deviations centered on each characteristic gamma ray energy. The standard deviations were calculated by dividing the photopeak FWHM obtained from the power-law fit in Figure~\ref{fig:energy_resolution} by the conversion factor (2.355). An energy window of $\pm 5$ standard deviations was chosen to accommodate potential energy calibration drift, the reduced energy resolution due to shorter acquisition time, and to ensure the inclusion of the quadratic background in the fitting region. The background-corrected radionuclide counts $\textbf{y}_{\text{c}}$ were obtained from the counts within $\pm 3$ standard deviations of the fitted Gaussian centroid(s). Figure~\ref{fig:at_altitude_radiation_map} displays heatmaps of background-corrected radionuclide counts from the live radioactive source demonstration.

\unskip

\begin{figure}[h!]
\includegraphics[width=0.495\linewidth]{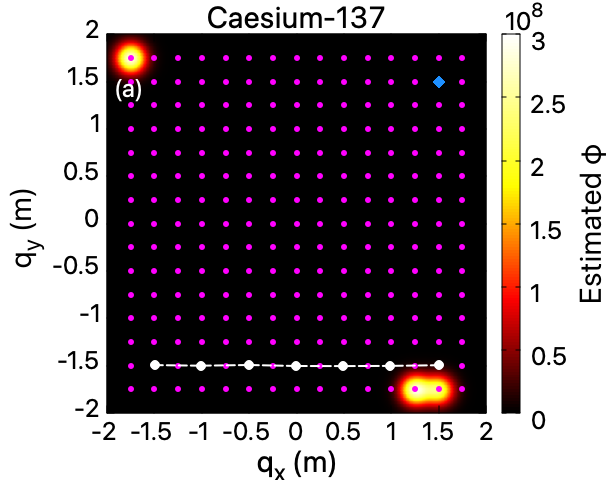}
\includegraphics[width=0.495\linewidth]{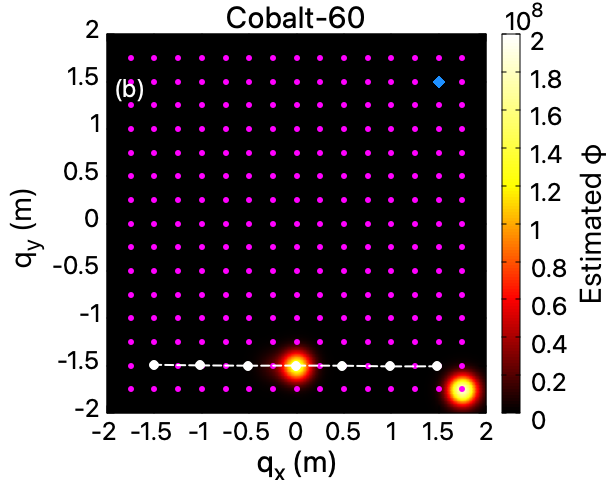}
\includegraphics[width=0.495\linewidth]{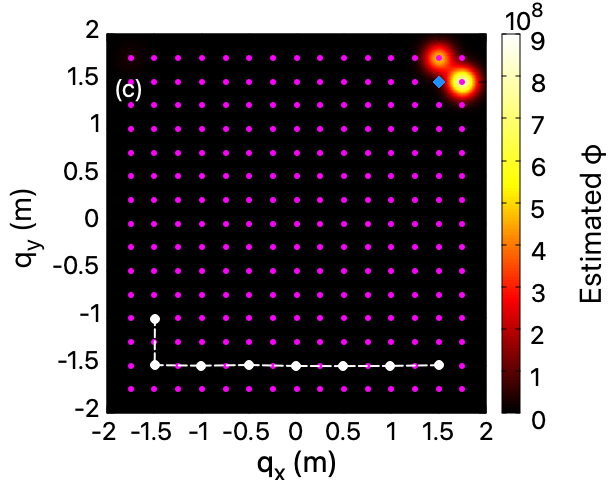}
\includegraphics[width=0.495\linewidth]{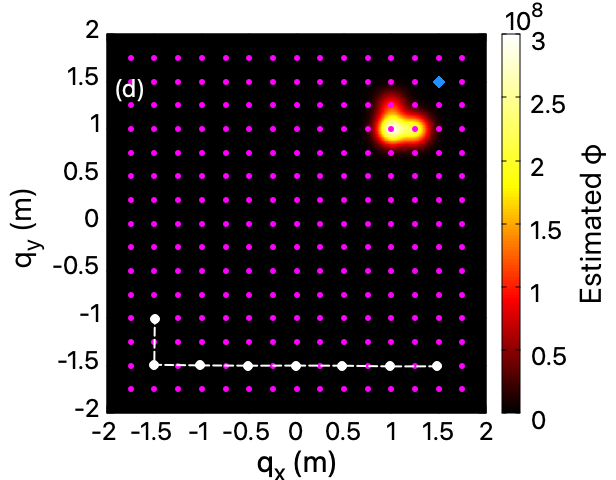}
\includegraphics[width=0.495\linewidth]{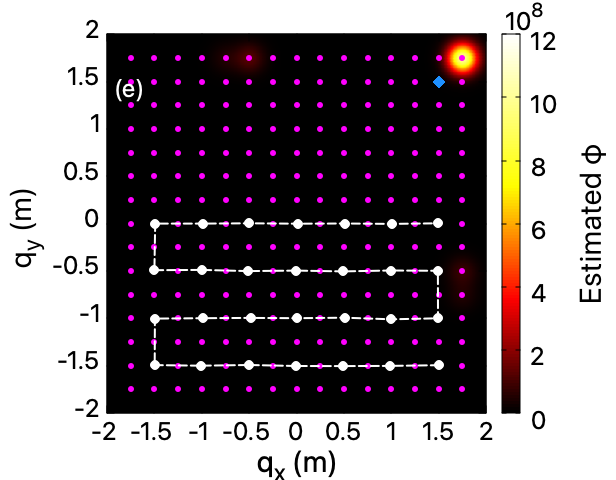}
\includegraphics[width=0.495\linewidth]{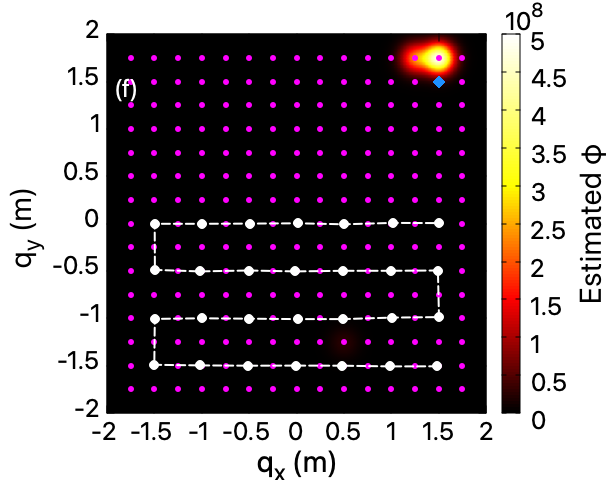}
\includegraphics[width=0.495\linewidth]{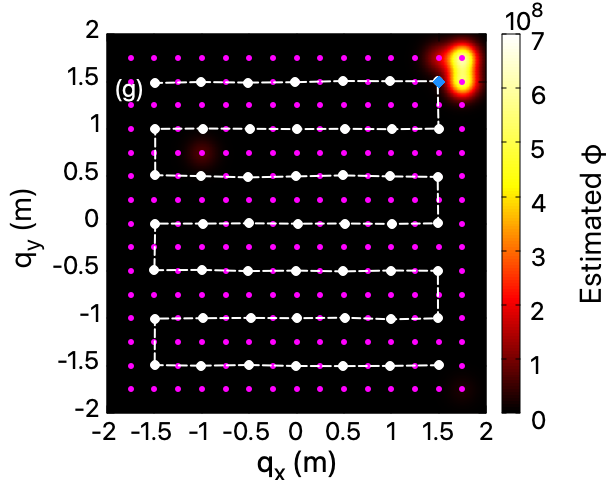}
\includegraphics[width=0.495\linewidth]{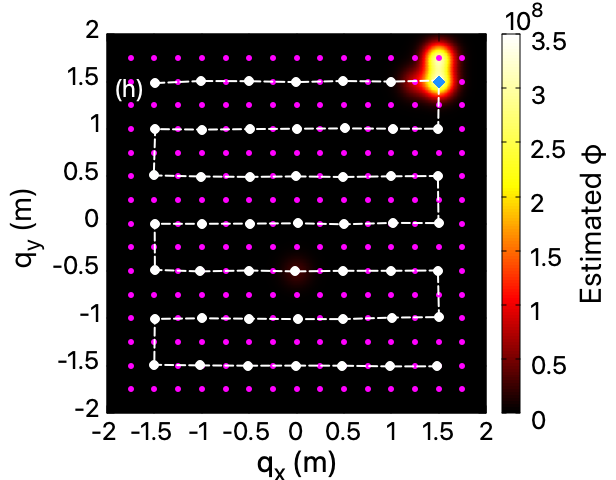}
\caption{Estimated ground-level source distributions from different time snapshots of radiation measurements (dotted-dashed line) along the 3 m-height UAV trajectory. The points indicate the grid of Dirac delta functions as the source distribution basis. Snapshots in (\textbf{a})-(\textbf{b}) and (\textbf{c})-(\textbf{d}) show the measurements positions before and after the initial resolution of the source location, (\textbf{e})-(\textbf{f}) illustrate the minimum number of measurements required to stabilise source localisation, and (\textbf{g})-(\textbf{h}) displays the complete raster scan. The flight trials with the $^{137}$Cs and $^{60}$Co sources (diamond marker at $x=1.5$ m and $y=1.5$ m) are shown in the left and right columns.} \label{fig:lls_rad_map_density_func_est}
\end{figure}

The source distribution in Equation~\ref{eq:dirac_decomposition_phi} was decomposed into a 3.5 m square grid consisting of 15 x 15 Dirac delta functions separated by 0.25 m along the ground plane. The matrix elements in Equation~\ref{eq:source_activity_estimation_algorithm} were computed for each measurement position and Dirac delta function combination, and subsequently used in Equation~\ref{eq:nnls_optimisation} to estimate the optimal basis weights $\hat{\boldsymbol{\beta}}$. Detector measurement positions were derived from the motion capture-measured platform position with a constant vertical offset of $z=-0.2$ m to account for the displacement of the NaIL scintillation detector relative to the center of the reflective markers, as shown in Figure~\ref{fig:payload_with_drone}. The platform attitude variation was negligible during measurements, with an average angular deviation of $1.90 \pm 0.01$ degrees from the flight level. Figure~\ref{fig:lls_rad_map_density_func_est} plots the estimated source distribution of $^{137}$Cs and $^{60}$Co at ground level for different time snapshots along the 3 m-height UAV trajectory. For ease of visualisation, each Dirac delta function was blurred in the heat plot using a Gaussian function with a standard deviation chosen so that neighbouring peaks overlapped at half their maximum height. Figures~\ref{fig:lls_rad_map_density_func_est}a-\ref{fig:lls_rad_map_density_func_est}b and ~\ref{fig:lls_rad_map_density_func_est}c-\ref{fig:lls_rad_map_density_func_est}d show the heat map before and after the initial localisation of the $^{137}$Cs and $^{60}$Co source locations. With additional measurements, the estimated source distribution fluctuated until the fourth pass of the raster scan where it began to stabilise (Figures~\ref{fig:lls_rad_map_density_func_est}e and \ref{fig:lls_rad_map_density_func_est}f). Finally, Figures~\ref{fig:lls_rad_map_density_func_est}g and \ref{fig:lls_rad_map_density_func_est}h show the converged source distributions following the completion of the full raster scan. Table~\ref{table:linear_least_squares_source_activity_estimate} presents the estimated activities of the $^{137}$Cs and $^{60}$Co sources, which were obtained by integrating the estimated source distributions over the ground plane.

\unskip

\begin{table}[h!]
\caption{$^{137}$Cs and $^{60}$Co source activity estimates. The error relative to the true activity is presented in brackets.}\label{table:linear_least_squares_source_activity_estimate}
\centering
\setlength{\tabcolsep}{10pt}
\begin{tabular}{|c||c|c|}
\hline
Flight height & $^{137}$Cs (MBq) & $^{60}$Co (MBq) \\
\hline
1.5 m & 84 (13\%) & 44 (10\%) \\
\hline
3 m & 92 (5\%) & 49 ($<1$\%) \\
\hline
4.5 m & 94 (3\%) & 52 (6\%) \\
\hline
\hline
True activity & 97 & 49 \\
\hline
\end{tabular}
\end{table}

\section{Discussion}\label{sec:discussion}
This work demonstrates the effectiveness of low-cost SiPM-based NaIL scintillation detectors for aerial radiation monitoring. The developed UAV-based radiation detection system was deployed to localise and estimate the activities of $^{137}$Cs and $^{60}$Co sources. Across all live radioactive source demonstration flights, the positions of the $^{137}$Cs and $^{60}$Co sources were estimated within 0.5 m of their true locations. Moreover, the source activities were estimated with errors on the order of 10\% or less. Having demonstrated the viability of this first-generation system for aerial radiation monitoring, future developments can focus on detector-driven navigation that dynamically adjusts the flight path of the system in response to live radiation measurements, rather than following preplanned waypoints.

The live radioactive source demonstration provided three key insights that will inform the future implementation of detector-driven navigation in the UAV-based radiation detection system. These insights were:
\begin{enumerate}
    \item introducing spatial excitation through range and angular variation to help capture more information about the environment (known as persistent excitation \cite{Narendra1984}),
    \item maintaining a minimum flight altitude to mitigate ground-induced Compton scattering, and
    \item operating at altitudes with adequate count rates.
\end{enumerate}
Insight (1) is demonstrated in Figure~\ref{fig:lls_rad_map_density_func_est}, which shows how the source localisation accuracy improved as additional measurements were collected. In the first pass of the raster scan, which comprised of seven measurements, the data was confined to a single spatial dimension, resulting in poor localisation accuracy (see Figures~\ref{fig:lls_rad_map_density_func_est}a and \ref{fig:lls_rad_map_density_func_est}b). The source localisation accuracy improved when the next measurement was recorded as it increased the spatial dimensionality of the data, as shown in Figures~\ref{fig:lls_rad_map_density_func_est}c and \ref{fig:lls_rad_map_density_func_est}d. A common approach to incorporating spatial excitation is to superimpose a dithered signal onto the trajectory of the platform \cite{Schwager2008}.

Insights (2) and (3) are demonstrated through Table~\ref{table:linear_least_squares_source_activity_estimate}, which shows the dependence of source activity estimate error on flight height. The relatively high errors of 13\% and 10\% at 1.5 m flight heights for $^{137}$Cs and $^{60}$Co were attributed to an increased detection of Compton-scattered gamma rays from the ground due to closer proximity. The error for $^{137}$Cs decreased to 5\% at 3 m and further to 3\% at 4.5 m. For $^{60}$Co, the uncertainty was less than 1\% at 3 m but increased to 6\% at 4.5 m. These results suggest the existence of an optimal operating altitude, dependent on the radionuclide count, that minimises the error in the source activity estimate. Operating above this altitude leads to sparser photopeak counts, causing increased estimation error. The optimal altitude has not been exceeded for the $^{137}$Cs flights, which explains the continuing decrease in the activity estimation error with altitude, whereas the maximum altitude has been surpassed for the $^{60}$Co flights, causing the activity error to increase at 4.5 m. This is consistent with Figures~\ref{fig:at_altitude_radiation_map}e and \ref{fig:at_altitude_radiation_map}f where the maximum radionuclide count for $^{137}$Cs is greater than $^{60}$Co at 4.5 m flight height.

\section{Conclusion}\label{sec:conclusion}
A UAV-based dual-mode gamma ray and neutron detection system for ground-level radioactivity estimation was developed, characterised, and demonstrated in flight for source localisation and activity
estimation applications. To support these capabilities, an experimentally validated radionuclide detection efficiency model was derived. The UAV-based radiation detection system and efficiency model were demonstrated through flight tests in a controlled environment, successfully localising $^{137}$Cs and $^{60}$Co sources within 0.5 m of their true positions and estimating their activities with errors on the order of 10\% or less. This study demonstrates the effectiveness of low-cost SiPM-based NaIL scintillation detectors for aerial radiation monitoring applications, such as homeland security, nuclear decontamination, and NORM detection.

\section*{Acknowledgment}
The authors would like to acknowledge the support of the Defence Science and Technology (DST) Group of the Australian Department of Defence in facilitating the flight trial. We thank Stephen Hardiman for his support in the early design of the UAV-based radiation detection system.

\bibliographystyle{IEEEtran}
\bibliography{export}

\end{document}